\documentclass[aps,prl,amsmath,amssymb,twocolumn]{revtex4-2}
\usepackage[utf8]{inputenc}
\usepackage{xcolor}
\definecolor{SPPBlue}{RGB}{56, 133, 194}
\usepackage[colorlinks=true,allcolors=SPPBlue,breaklinks=true]{hyperref}

\makeatletter
\newcommand*\SMTOCoff{\let\SM@l@section\l@section \let\l@section\@gobbletwo}
\newcommand*\SMTOCon{\let\l@section\SM@l@section}
\makeatother

\usepackage{xurl}
\usepackage{titlesec}
\usepackage{etoolbox}
\usepackage{graphicx}
\usepackage[normalem]{ulem}
\usepackage{amsmath}
\usepackage{amssymb}
\usepackage{bm}
\usepackage{mathtools}
\usepackage{braket}
\usepackage{float}
\usepackage{orcidlink}
\usepackage{balance}

\allowdisplaybreaks

\begin{document}

\title{
Birefringent Curvature Control of Surface Plasmons and Collective Emission
}

\author{Florian B\"onsel\,\orcidlink{0000-0002-7193-9643}}
\email[]{florian.boensel@mpl.mpg.de}
\affiliation{Max Planck Institute for the Science of Light, 91058 Erlangen, Germany}
\affiliation{Department of Physics, Friedrich-Alexander-Universit\"at Erlangen-N\"urnberg, 91058 Erlangen, Germany}

\author{Flore K.~Kunst\,\orcidlink{0000-0003-0445-0036}}
\email[]{flore.kunst@mpl.mpg.de}
\affiliation{Max Planck Institute for the Science of Light, 91058 Erlangen, Germany}
\affiliation{Department of Physics, Friedrich-Alexander-Universit\"at Erlangen-N\"urnberg, 91058 Erlangen, Germany}

\begin{abstract}
We present a generic wave equation for surface plasmon polaritons on any macroscopically curved metal-dielectric interface, with isotropic and anisotropic geometric potentials linear in curvature. Remarkably, the anisotropic birefringence vanishes if the metal-to-dielectric permittivity ratio equals the golden ratio squared, mimicking isotropy at linear order on an actually anisotropic surface, as confirmed by full-wave simulations. Applying our equation to quantum emitters on metallic spheres, we demonstrate a curvature-controlled reshaping of collective decay rates and frequency shifts.
\end{abstract}

\maketitle

\textit{Introduction.}---Curved surfaces reshape the waves traveling along them. Whether it matters which way a surface bends depends on how the mode is distributed across it. For light in a dielectric slab and related surface-guided waves~\cite{batz2008linear, schultheiss2020light, carmi2025photon, batz2010solitons, willatzen2009electromagnetic, schultheiss2010optics, zhang2024diffraction}, the modes decay symmetrically across the guiding layer. In this setting, curvature enters the geometric potentials only at second order---as shown for quantum particles in thin layers~\cite{da1981quantum, ferrari2008schrodinger, mazharimousavi2025quantum}, electronic materials~\cite{gentile2022electronic, marchi2005coherent}, and Bose-Einstein condensates~\cite{salasnich2022bose, sandin2017dimensional}---so that convex and concave bending are indistinguishable.

\begin{figure}[H]
\includegraphics[width=\columnwidth]{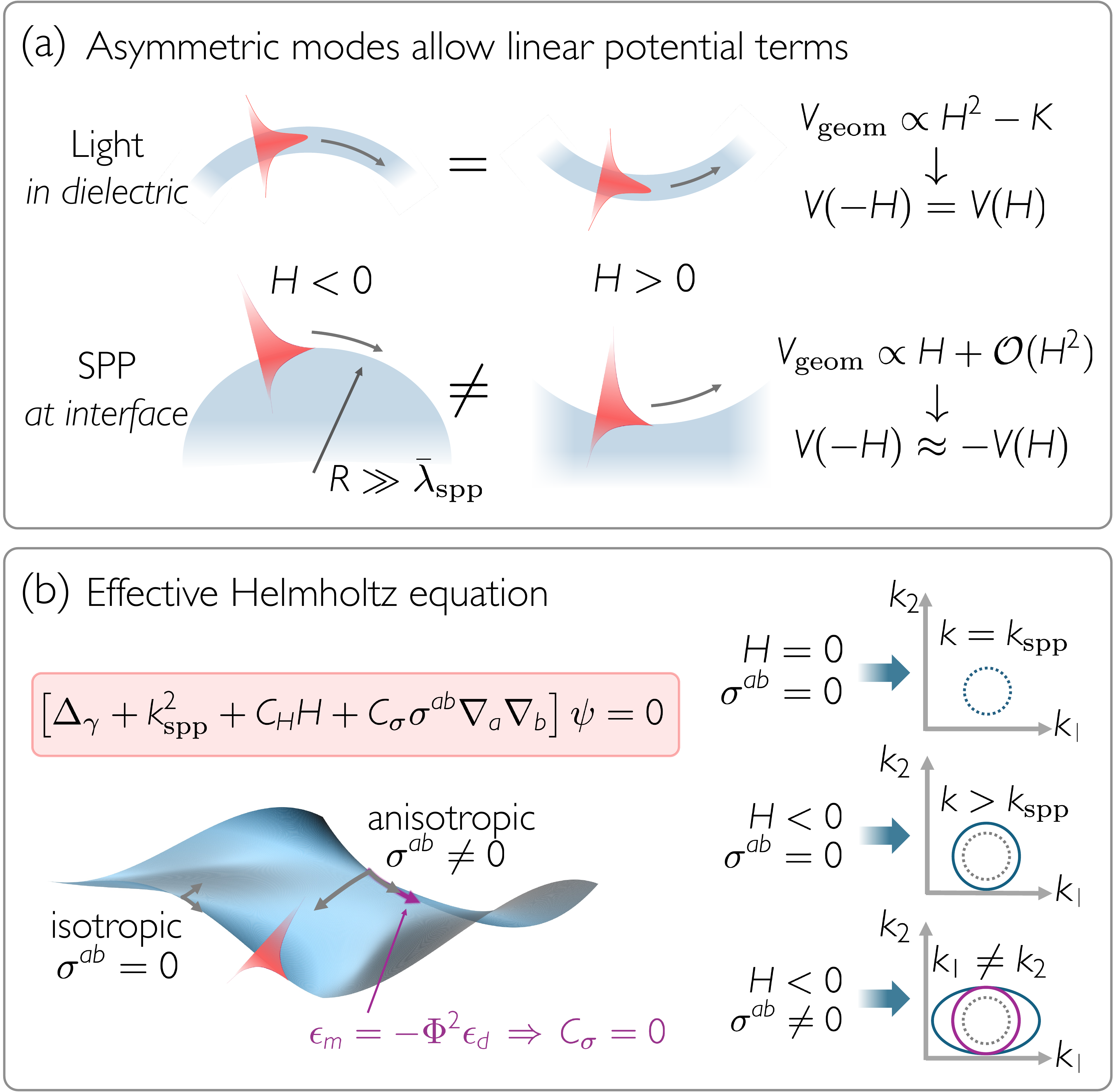}
\caption{\textit{Surface plasmon polaritons on curved interfaces.} (a)~Symmetric dielectric confinement yields a curvature-quadratic geometric potential (top), while the asymmetric SPP modes give a linear one (bottom). (b)~Wave equation for the SPP envelope $\psi$. Right: isotropic curvature rescales the flat wavevector circle (blue solid, middle), anisotropic curvature turns it into an ellipse (bottom), circular again at the metal-dielectric permittivity relation $\epsilon_m=-\Phi^2\epsilon_d$ with $\Phi$ the golden ratio (purple solid). Symbols: $R$: local curvature radius; $\lambda_{\rm spp}$: SPP wavelength; $H$ and $K$: extrinsic and intrinsic curvature, respectively; $\Delta_\gamma$: Laplace-Beltrami operator; $k_{\rm spp}$: flat-interface wavenumber; $\sigma^{ab}$: traceless second fundamental form; $C_{H,\sigma}$: material prefactors.}
\label{fig:introduction_image}
\end{figure}

Surface plasmon polaritons (SPPs) are different. As modes existing \textit{at an interface} between a metallic and dielectric material rather than primarily \textit{inside a material}, their field decays more strongly into the metal than into the dielectric. We show that this asymmetric confinement makes convex and concave bending distinguishable, with geometric corrections appearing already at first order, cf.~Fig.~\ref{fig:introduction_image}~(a). Curvature therefore becomes a sign-sensitive handle on these modes and on the interactions they mediate, which motivates this Letter.

Confining light to the interface below the diffraction limit, SPPs can mediate strong light-matter interactions~\cite{torma2014strong, chang2007strong, gonzalez2013theory}, and underlie a broad range of lattice and topological effects, mostly on discrete patterned structures~\cite{kravets2018plasmonic, guo2017geometry, weick2013dirac, jia2024cascade, freire2025plasmonic, xiong2019photonic}, while nanofabrication delivers smooth metallic interfaces with prescribed out-of-plane curvature~\cite{lassaline2020optical, nagpal2009ultrasmooth}. For cylindrical geometries, a sign-dependent geometric potential for curvature-induced waveguiding was predicted as a paraxial Schr\"odinger-type equation~\cite{della2010geometric}, confirmed experimentally~\cite{libster2019surface}, and used to study a one-dimensional topological plasmonic lattice~\cite{smith2021topological}. Further studies of cylinders and spheres revealed curvature-modified dispersions and directional momentum corrections~\cite{ancey2009surface, liaw2008dispersion, spittel2015curvature, perel2011asymptotics, hasegawa2007curvature, vasconcelos1991surface}. These treatments are paraxial, ray-based, or restricted to specific surface trajectories. No governing equation exists for the full two-dimensional dynamics on an arbitrarily shaped interface.

\begin{figure}
\includegraphics[width=\columnwidth]{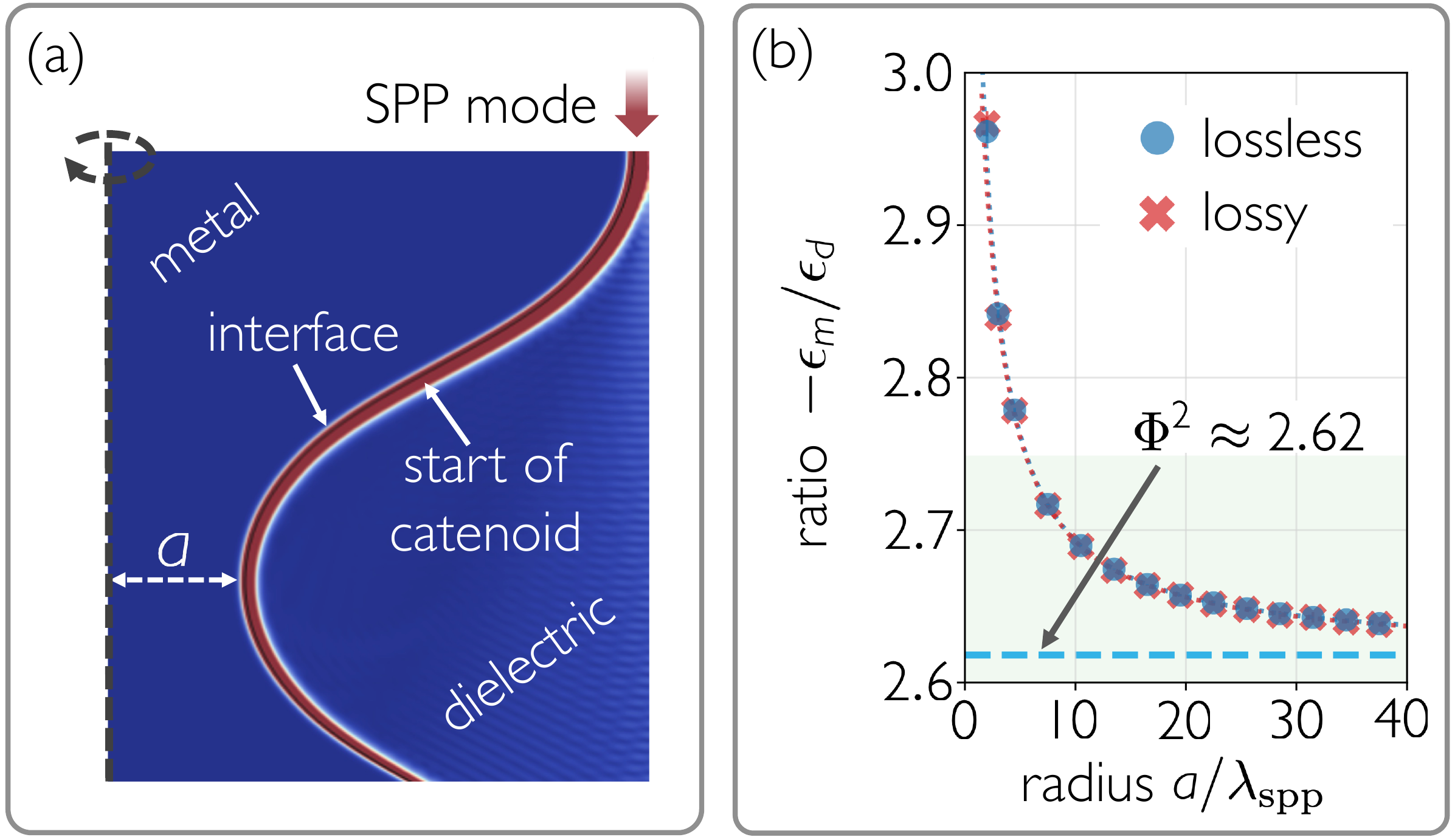}
\caption{\textit{Golden-ratio prediction tested on a metal catenoid.} (a)~Cross section of the axisymmetric setup, with an SPP launched on a lossy metal cylinder (top) and propagating smoothly into the catenoid section of waist radius $a$. Left: metal, right: dielectric. Color shows the electric field strength increasing from blue to red. (b)~Ratio $-\epsilon_m/\epsilon_d$ at which the curvature-induced birefringence vanishes, versus $a/\lambda_{\rm spp}$. It converges to $\Phi^2\approx 2.62$ (dashed horizontal) for $a\gg\bar{\lambda}_{\rm spp}$. Dotted curves: Quadratic fit in $\lambda_{\rm spp}/a$, green shaded: $5\,\%$ window around $\Phi^2$. Both panels: $\lambda_{\rm spp}=230\,\mathrm{nm}$, $\epsilon_d=2.25$.}
\label{fig:catenoid}
\end{figure}

In this Letter, we fill this gap. From Maxwell's equations with curvature-modified boundary matching conditions, we obtain an effective two-dimensional Helmholtz equation for the in-plane SPP mode on a surface with macroscopic curvature, i.e., a curvature radius large com-
\noindent
pared to the SPP wavelength. The Helmholtz equation  includes two potentials linear in curvature: an isotropic one set by the extrinsic curvature, and an anisotropic one set by the traceless second fundamental form, which acts as a curvature-induced SPP birefringence, cf.~Fig.~\ref{fig:introduction_image}~(b). Not only does our equation recover known results for spheres and cylinders~\cite{della2010geometric,smith2021topological,ancey2009surface,spittel2015curvature,perel2011asymptotics}, 
it has two further consequences. First, we predict that the anisotropic term vanishes at a metal-to-dielectric permittivity ratio equal to the golden ratio squared, rendering a surface isotropic to the SPP, as full-wave simulations of a metal catenoid confirm, cf.~Fig.~\ref{fig:catenoid}. Second, Ohmic damping becomes curvature dependent, so that the shape of an interface also determines the non-Hermitian structure of the mode.

\begin{figure}
\includegraphics[width=\columnwidth]{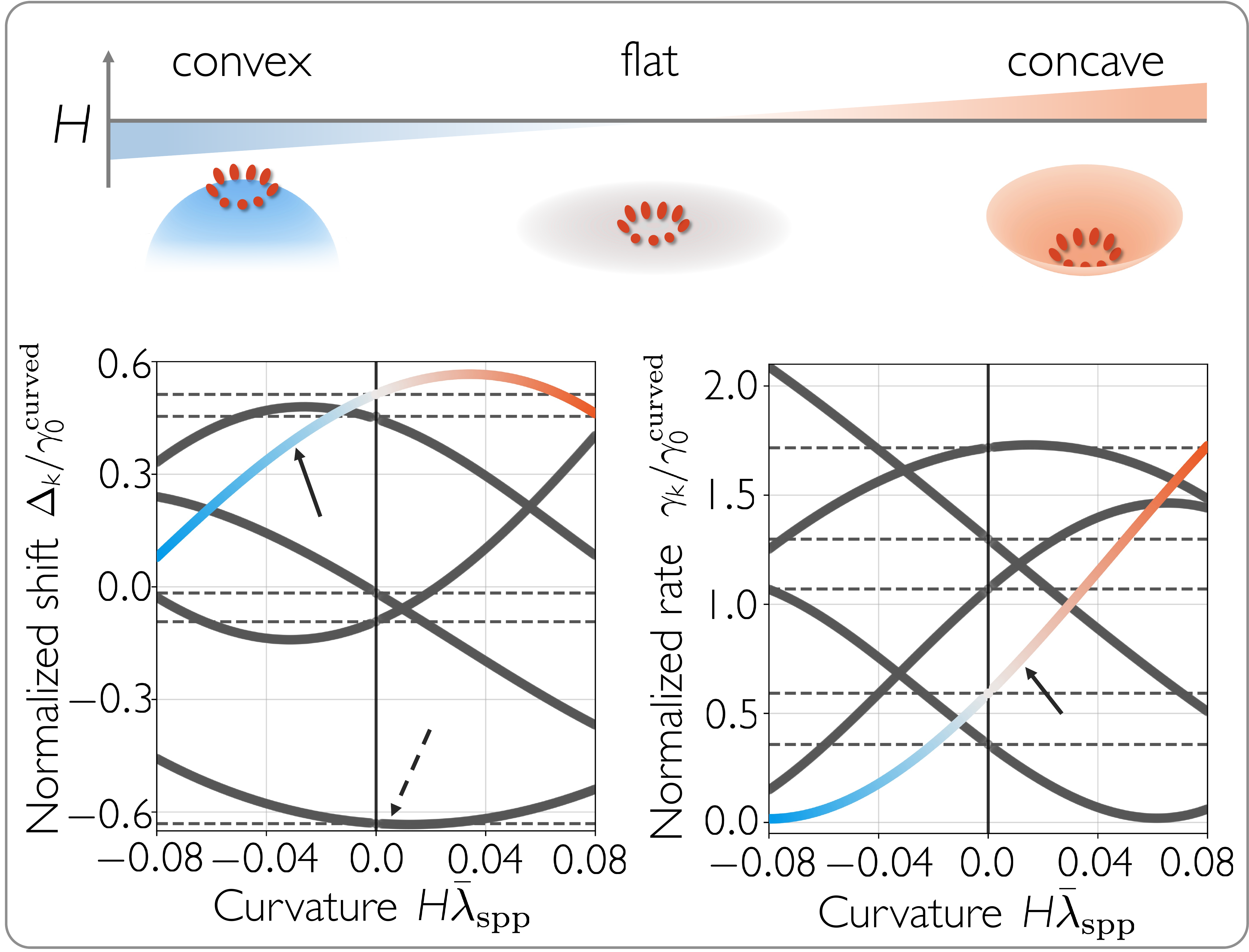}
\caption{\textit{Collective emitter radiance under curvature.} Bottom panel: Normalized collective frequency shifts (left) and decay rates (right) for a ring of $N=9$ emitters on the pole of a silver sphere in air at $\lambda_0=600\,$nm, versus curvature $H\bar{\lambda}_{\rm spp}$: convex ($H<0$, blue) through flat (gray) to concave ($H>0$, red), see top panel. Emitter height and geodesic spacing ($3\bar{\lambda}_{\rm spp}$) are fixed as the sphere is deformed.  Normalizing by the single-emitter rate $\gamma_0^{\rm curved}$ on the same surface divides out the Purcell change. Dashed lines: planar reference; black: all modes. Arrows mark the mode highlighted by a color gradient (solid) and a mode responding weakly to curvature (dashed).}
\label{fig:cooperativity}
\end{figure}

As an application, we use the Helmholtz equation to study SPP-mediated dipole coupling between emitters placed on a metallic sphere. While cooperative decay rates and frequency shifts of coupled emitters via SPPs are well studied near planar 
interfaces~\cite{novotny2012principles,torma2014strong, chance1978molecular, barnes1998fluorescence, jalali2024collective, jorgensen2025collective}
and nanowires~\cite{chang2006quantum, dzsotjan2010quantum,dzsotjan2011dipole, gonzalez2011entanglement, pustovit2010plasmon, barthes2013coupling}, for which curvature is either absent or so tight that the mode is squeezed into the subwavelength regime, the effect of macroscopic curvature has so far been neglected. This regime is interesting because it allows SPPs to propagate across extended emitter arrangements, and a linear potential should imprint the sign of the bending onto the collective spectrum. Through a Green's function derived from the wave equation, we indeed find that curvature redistributes the modes between super- and subradiant channels asymmetrically in the sign of the bending, cf.~Fig.~\ref{fig:cooperativity}.

\textit{Curved-surface Helmholtz equation.}---On a flat interface, a transverse-magnetic SPP mode exists provided $\epsilon_m < -\epsilon_d$, with $\epsilon_m$ and $\epsilon_d$ the metal and dielectric permittivities, respectively. We take non-magnetic materials, and real $\epsilon_d$, a very good approximation in the visible regime~\cite{gao2013refractive}, while keeping $\epsilon_m$ real for now (we consider Ohmic loss below). The SPP dispersion relation and inverse decay lengths read
\begin{equation}\label{eq:kspp_kappa}
k_{\mathrm{spp}} = k_0\sqrt{\frac{\epsilon_m\epsilon_d}{\epsilon_m+\epsilon_d}}, \qquad \kappa_{d,m} = \sqrt{k_{\mathrm{spp}}^2-\epsilon_{d,m}k_0^2},
\end{equation}
respectively, with $k_0$ the vacuum wavenumber. They obey $\kappa_d\kappa_m = k_{\mathrm{spp}}^2$, so a single scale fixes the mode structure both along and across the interface. Working with the electromagnetic four-potential in the Lorenz gauge, the SPP carries a single nonzero spatial component along the interface normal $\eta$, whose positive direction points into the dielectric. It separates as $e^{-\kappa_{d,m}|\eta|}\psi$ into a profile across the interface, and an in-plane envelope $\psi$ obeying $[\Delta+k_{\mathrm{spp}}^2]\psi=0$. Every field component follows from $\psi$, so the flat SPP is a two-dimensional Helmholtz problem.

Let us now extend this to the curved regime. The local shape of a curved interface is encoded in the second fundamental form $h_{ab} = H\gamma_{ab} + \sigma_{ab}$, with $\gamma_{ab}$ the intrinsic surface metric, $H = (\kappa_1+\kappa_2)/2$ the extrinsic curvature as the mean of the two principal curvatures $\kappa_{1,2}$, and $\sigma_{ab}$ the traceless remainder. We adopt the convention that a convex metal surface has $H<0$. Curvature enters through the ratio $\alpha = |H|/k_{\mathrm{spp}}\sim\bar{\lambda}_{\mathrm{spp}}/R$ of the reduced SPP wavelength $\bar{\lambda}_{\mathrm{spp}}$ to the local curvature radius $R$, which is our small expansion parameter in the macroscopic-curvature regime of interest. We further assume that the curvature itself varies slowly on the SPP scale, i.e., $|\nabla_a H| \ll k_{\rm spp}|H|$.

Solving Maxwell's equations order by order in $\alpha$ with curvature-modified boundary conditions, we obtain the effective two-dimensional wave equation for the in-plane SPP envelope on the curved interface as
\begin{equation}\label{eq:main_equation}
\left[\Delta_\gamma + k_{\mathrm{spp}}^2 + V_H + V_\sigma\right]\psi = 0,
\end{equation}
with $\Delta_\gamma$ the Laplace-Beltrami operator, an isotropic scalar potential $V_H = C_HH$, and an anisotropic operator $V_\sigma = C_\sigma\sigma^{ab}\nabla_a\nabla_b$, with material-dependent coefficients
\begin{subequations}
\begin{align}
C_H &= \frac{k_0\left(\epsilon_d^2+\epsilon_d\epsilon_m+\epsilon_m^2\right)}{\left(\epsilon_d+\epsilon_m\right)\sqrt{-\left(\epsilon_d+\epsilon_m\right)}},\label{eq:C_H}\\
C_\sigma &= -\frac{\epsilon_d^2+3\epsilon_d\epsilon_m+\epsilon_m^2}{k_0\,\epsilon_d\epsilon_m\sqrt{-\left(\epsilon_d+\epsilon_m\right)}}.\label{eq:C_sigma}
\end{align}
\end{subequations}
Equation~\eqref{eq:main_equation} is the central result of this work, and is accurate to first order in curvature and valid on any macroscopically curved metal-dielectric interface, cf.~Fig.~\ref{fig:introduction_image}~(b). The derivation of this equation is outlined in the End Matter, and all technical steps are provided in the Supplemental Material (SM)~\cite{supp}. Since the intrinsic geometry, entering through the Gaussian curvature $K$, departs from its flat counterpart at $\mathcal{O}(\alpha^2)$, flat derivative operators suffice at $\mathcal{O}(\alpha)$. Writing them covariantly in Eq.~\eqref{eq:main_equation} leaves this order untouched, while making the result applicable to closed surfaces without patching local charts.

Both potentials are linear in the curvature, distinguishing SPPs from waves guided inside a material, where the leading geometric potential is quadratic~\cite{da1981quantum, ferrari2008schrodinger, batz2008linear, carmi2025photon}. The reason is that the SPP samples the geometry at a finite distance from the interface, where the metric deviates from its interface value by $\sim\eta H$ with opposite signs on the two sides of it. For a symmetrically confined mode these deviations cancel, leaving a response only at $\mathcal{O}(\alpha^2)$, cf.~Fig.~\ref{fig:introduction_image}~(a), whereas the SPP reaches $\kappa_d^{-1}$ in the dielectric but only $\kappa_m^{-1}<\kappa_d^{-1}$ in the metal, so that a term linear in the curvature survives. Retaining a finite $\eta$ is therefore essential here, unlike in the Jensen-Koppe-da-Costa thin-layer limit $\eta\to0$~\cite{jensen1971quantum, da1981quantum} frequently applied to light in curved thin dielectrics~\cite{batz2008linear, carmi2025photon, longhi2007topological, schultheiss2020light}.

Let us make a few remarks. Firstly, the two potentials enter the local dispersion differently. A local plane wave ansatz $\psi\sim e^{ik_aq^a}$, with $q^a$ the in-plane surface coordinates, turns Eq.~\eqref{eq:main_equation} into
\begin{equation}\label{eq:local_dispersion}
\left(1+C_\sigma S\right)k^2 = k_{\mathrm{spp}}^2 + C_HH, \qquad S\equiv\sigma^{ab}n_an_b,
\end{equation}
with $n_a=k_a/k$ the propagation direction. Since $C_H<0$, convex bending ($H<0$) shortens the SPP wavelength, while concave bending ($H>0$) stretches it, rescaling the flat wavevector circle, cf.~Fig.~\ref{fig:introduction_image}~(b). The anisotropic term instead deforms it into an ellipse, as $S=\pm(\kappa_1-\kappa_2)/2$ along the principal directions, and vanishes where these coincide, as on a sphere. Both $H$ and $\sigma^{ab}$ vary across a general interface, so the dispersion turns direction and position dependent. This behavior is different from bent dielectric waveguides, which show a birefringence quadratic in curvature that splits two polarizations~\cite{ulrich1980bending, smith1980birefringence}. Secondly, in the End Matter, we show that our result captures the known dispersion and paraxial propagation results for spheres~\cite{ancey2009surface, perel2011asymptotics} and cylinders~\cite{della2010geometric, spittel2015curvature}.

\textit{Curvature and Ohmic loss.}---For real $\epsilon_m$, the operator in Eq.~\eqref{eq:main_equation} is Hermitian at $O(\alpha)$. Due to Ohmic loss, realistic metals have complex permittivities $\epsilon_m+i\epsilon_m'$, however, with $0<\epsilon_m'\ll|\epsilon_m|$ at optical frequencies, which we treat perturbatively. Expanding Eq.~\eqref{eq:kspp_kappa} to first order in $\epsilon_m'$ gives $k_{\rm spp}^2\to k_{\rm spp}^2+iK_{\rm loss}$, with $K_{\rm loss}=k_0^2\epsilon_d^2\epsilon_m'/(\epsilon_d+\epsilon_m)^2$ a spatially uniform Ohmic damping. The coefficients $C_H$ and $C_\sigma$ expand in the same way as $C_{H,\,\sigma}\to C_{H,\,\sigma}+iC_{H,\,\sigma}'$, with imaginary parts proportional to $\epsilon_m'$, so that Eq.~\eqref{eq:main_equation} becomes
\begin{equation}\label{eq:lossy_equation}
\begin{aligned}
    &\left[\Delta_\gamma + k_{\text{spp}}^2 + C_H H + C_\sigma\,\sigma^{ab}\nabla_a\nabla_b\right]\psi \\
    &\quad + \,i\left[K_{\text{loss}} + C_H' H + C_\sigma'\,\sigma^{ab}\nabla_a\nabla_b\right]\psi = 0.
\end{aligned}
\end{equation}
The terms inside the first bracket reproduce the lossless Eq.~\eqref{eq:main_equation}, while the terms inside the second bracket govern dissipation. A uniform $K_{\rm loss}$ causes only an overall decay and leaves the mode structure untouched, whereas the curvature terms tie the damping to the shape of the interface, through $C_H'H$ to the sign of the bending and through $C_\sigma'S$ to the propagation direction, so that shaping an interface is what gives the SPP a non-trivial non-Hermitian structure~\cite{ashida2020non, Bergholtz2021}.

\textit{Golden ratio.}---The anisotropy coefficient $C_\sigma$ in Eq.~\eqref{eq:C_sigma} changes sign when its numerator vanishes,
\begin{equation}\label{eq:golden_ratio_condition}
\epsilon_d^2+3\epsilon_d\epsilon_m+\epsilon_m^2 = 0 \quad\Longleftrightarrow\quad \epsilon_m = -\Phi^2\epsilon_d,
\end{equation}
with $\Phi = (1+\sqrt{5})/2$ the golden ratio, while its second root $-\Phi^{-2}\epsilon_d$ violates the SPP existence condition $\epsilon_m<-\epsilon_d$. There, Eq.~\eqref{eq:main_equation} collapses to $[\Delta_\gamma+k_{\mathrm{spp}}^2+C_HH]\psi = 0$ and the wavevector ellipse of Eq.~\eqref{eq:local_dispersion} degenerates into a circle, so that the SPP responds to an actually anisotropic surface as if it were isotropic. The isotropic potential is untouched, since the numerator of $C_H$ has no real roots, so that convex and concave bending stay distinguishable at this point. An equivalent form, $(\epsilon_m+\epsilon_d)^2 = -\epsilon_m\epsilon_d$, was mentioned in Ref.~\cite{spittel2015curvature} for a metal wire, whereas we find it to hold on any weakly curved surface.

This prediction is robust under dissipation. Expanding Eq.~\eqref{eq:local_dispersion} at $H=0$ gives $k\approx k_{\rm spp}(1-C_\sigma S/2)$, so that the curvature-induced phase follows $\mathrm{Re}[k_{\rm spp}C_\sigma]$ and the isotropy point is where this combination vanishes. Both factors acquire imaginary parts of first order in $\epsilon'_m$, whose product displaces it only at second order as
\begin{equation}\label{eq:golden_ratio_loss}
\frac{\epsilon_m}{\epsilon_d} = -\Phi^2 + \frac{5+\sqrt5}{20}\left(\frac{\epsilon'_m}{\epsilon_d}\right)^2,
\end{equation}
amounting to a relative correction of $0.03\,\%$ for the $\epsilon'_m=0.1$ and $\epsilon_d=2.25$ used below.

To test both predictions, we simulate SPP propagation on a catenoid, a minimal surface with $H=0$, which isolates $V_\sigma$. Since $C_\sigma$ changes sign at Eq.~\eqref{eq:golden_ratio_condition}, so does the phase accumulated relative to a flat interface of equal length, and the condition shows up as a zero crossing when the permittivity ratio is swept. In full-wave simulations~\footnote{For the full-wave simulations, we employed commercial software (COMSOL Multiphysics, version~6.4).}, we launch a fundamental mode SPP on a cylinder that bends smoothly into a catenoid of waist radius $a=2\,\lambda_{\mathrm{spp}}$--$38.5\,\lambda_{\mathrm{spp}}$, cf.~Fig.~\ref{fig:catenoid}~(a), and locate that phase crossing at fixed waist. An experiment or simulation is limited by the SPP propagation length. As an estimate, the field on a flat and lossy metal varies as $\sim e^{i\left(k_{\rm spp}+iK_{\rm loss}/2k_{\rm spp}\right)x}$ with distance  $x$~\cite{maier2007plasmonics}, and decays by $1/e$ over $L_{\rm spp}=2k_{\rm spp}/K_{\rm loss}$, that is, over $L_{\rm spp}/\lambda_{\rm spp}=\epsilon_m(\epsilon_d+\epsilon_m)/(\pi\epsilon_d\epsilon'_m)$ SPP wavelengths. At the condition of Eq.~\eqref{eq:golden_ratio_condition}, this reads $L_{\rm spp}/\lambda_{\rm spp}=\Phi^3\epsilon_d/(\pi\epsilon'_m)$, so that at $\epsilon'_m=0.1$ the SPP travels $30\,\lambda_{\rm spp}$ for $\epsilon_d=2.25$, but only $13\,\lambda_{\rm spp}$ in air with $\epsilon_d=1$, which is too short to traverse the full catenoid section for large waists.

The extracted ratios lie remarkably close to the prediction, and they approach it as the waist grows and the neglected $O(\alpha^2)$ terms shrink, with the extrapolation to vanishing curvature giving $-\epsilon_m/\epsilon_d = \Phi^2\approx2.62$, see Fig.~\ref{fig:catenoid}~(b). Adding loss, i.e., $\epsilon_m\to\epsilon_m+i\epsilon'_m$ with $\epsilon'_m=0.1$, changes the extracted ratios by less than $0.2\,\%$ at all considered sizes, whereas a displacement of first order in $\epsilon'_m$ would amount to $\approx2\,\%$, which confirms the second-order scaling of Eq.~\eqref{eq:golden_ratio_loss}. Since Eq.~\eqref{eq:golden_ratio_condition} constrains only the permittivities, the isotropy point is reached by tuning the material combination rather than the shape, so that changing the dielectric environment sweeps an SPP through it on one and the same sample.

\textit{Collective emission.}---As an application of Eq.~\eqref{eq:lossy_equation} beyond bare SPP dynamics, we turn to SPP-mediated dipole coupling, which together with Purcell enhancement and cooperative decay is well understood near flat interfaces~\cite{zhou2011surface, zhou2017enhanced, mueller2013asymmetric, pockrand1980nonradiative, choquette2010superradiance, kundu2025cooperative, novotny2012principles, torma2014strong, jalali2024collective, jorgensen2025collective} and on sub-wavelength nanostructures~\cite{chang2007strong, gonzalez2013theory, stehle2014cooperative, jalali2026collective, liu2025quantum, bermudez2015coupling, chang2006quantum, dzsotjan2010quantum, dzsotjan2011dipole, gonzalez2011entanglement, pustovit2010plasmon, barthes2013coupling}. Within tens of nanometers above a metal plane, this channel dominates over the radiative and lossy contributions~\cite{chance1978molecular, barnes1998fluorescence}, a regime realizable by dye molecules, quantum dots, and nitrogen-vacancy centers~\cite{wei2011quantum, rose2014control, li2015quantum, huck2011controlled}. We ask how macroscopic curvature reshapes it, and place $N$ two-level emitters of transition frequency $\omega_0$, with dipole moments oriented normal to the surface, on a ring~\cite{scheil2023optical} near the pole of a metal sphere of radius $R$ at a fixed height $d$, with $\kappa_dd\ll1$. Since the principal curvatures coincide on a sphere, $\sigma^{ab}$ vanishes and $V_H$ is the only geometric potential entering here.

In the single-excitation sector and under the Markov approximation, the array is governed by the effective non-Hermitian Hamiltonian~\cite{asenjo2017exponential}
\begin{equation}
\mathcal{H} = \hbar\omega_0\sum_{i=1}^{N}\hat{\sigma}^i_{ee} + \hbar\sum_{i,j=1}^{N}\left(J^{ij}-i\frac{\Gamma^{ij}}{2}\right)\hat{\sigma}^i_{eg}\hat{\sigma}^j_{ge},
\end{equation}
with $\hbar$ the reduced Planck constant, $\hat{\sigma}^i_{ee}$ the excited-state population operator, and $\hat{\sigma}^i_{eg}=(\hat{\sigma}^i_{ge})^\dagger$ the transition operators of emitter $i$. Because the dipoles point along $\eta$, the coherent and dissipative rates $J^{ij}-i\Gamma^{ij}/2\propto G_{\eta\eta}(\mathbf{r}_{0,i},\mathbf{r}_{0,j})$ are set by the normal-normal component of the dyadic Green's tensor at the emitter positions $\mathbf{r}_{0,i}=(\mathbf{q}_{0,i},d)$. We obtain $G_{\eta\eta}$ from the same curved-surface equation, now sourced by a point emitter as
\begin{equation}\label{eq:greens_function}
\mathcal{L}\,G^{\mathrm{curved}}_{\mathrm{2D}}(\mathbf{q},\mathbf{q}_0) = -\delta^{(2)}(\mathbf{q}-\mathbf{q}_0)/\sqrt{\gamma},
\end{equation}
with $\mathcal{L}$ the operator acting on $\psi$ in Eq.~\eqref{eq:lossy_equation} and $\gamma=\det\gamma_{ab}$, and through the factorization $G_{\eta\eta}\approx C_0\,G^{\mathrm{curved}}_{\mathrm{2D}}$, where $C_0$ is a geometry-independent material constant. On a flat interface, the residue at the SPP pole of the transverse-magnetic Fresnel coefficient gives $G_{zz}=C_0e^{-2\kappa_dd}G^{\mathrm{flat}}_{\mathrm{2D}}$~\cite{novotny2012principles}, reflecting the separability $A^z=f(z)\psi(x,y)$. Curvature adds polynomial corrections in $\eta$ that reduce to a constant for $\kappa_dd\ll1$, so the factorization carries over (see the SM~\cite{supp}). Writing the $k$-th eigenvalue as $\lambda_k=\Delta_k-i\gamma_k/2$ and normalizing by the renormalized single-emitter rate $\gamma_0^{\mathrm{curved}}$, modes with $\gamma_k\gtrless\gamma_0^{\mathrm{curved}}$ are super- and subradiant relative to a single emitter on the same surface.

For $N=9$ emitters on a silver sphere in air ($\epsilon_m+i\epsilon'_m=-16.12+0.44i$ at $\lambda_0=600\,$nm~\cite{johnson1972optical}, $\bar{\lambda}_{\mathrm{spp}}=92.5\,$nm),  Fig.~\ref{fig:cooperativity} shows the normalized collective eigenvalues against $H\bar{\lambda}_{\mathrm{spp}}$, down to $R_{\mathrm{min}}=12.5\,\bar{\lambda}_{\mathrm{spp}}$. The spectrum is clearly asymmetric under $H\to-H$. Convex and concave curvatures redistribute decay rates and frequency shifts differently and mode-dependently. Some modes turn from nearly dark to superradiant on one side of $H=0$, while others behave the reversed way. The origin is the linear coupling of curvature to the effective wavenumber, $k^2_{\mathrm{eff}}=k^2_{\mathrm{spp}}+C_HH$. Under $H\to-H$, the dispersion shift changes sign, so the collective spectrum cannot be symmetric. Artificially setting $C_H=0$ removes the asymmetry, leaving only weak, symmetric corrections from the intrinsic Laplace-Beltrami operator (see Fig.~S1(a) in the SM~\cite{supp}), confirming that $V_H$ and not higher-order intrinsic-curvature terms drives the redistribution in the regime considered here. 

Interestingly, the modes further differ in how strongly they respond. The highlighted mode (colored, solid arrows in Fig.~\ref{fig:cooperativity}) shifts significantly away from $H=0$, while others (dashed arrow, left panel) barely change. Therefore, $V_H$ acts selectively on the collective spectrum and does not rescale it as a whole.

\textit{Conclusion \& Outlook.}---We derived a surface wave equation that holds on any macroscopically curved metal-dielectric interface and carries two geometric terms linear in curvature, an isotropic one and a birefringent one. It reproduces earlier geometry-specific results for the sphere and cylinder, as shown in the End Matter, it predicts a generic isotropy point at $\epsilon_m=-\Phi^2\epsilon_d$ that we confirm in full-wave simulations, and, applied to a ring of emitters on a metal sphere, it shows that curvature redistributes superradiant and subradiant channels asymmetrically in the sign of the bending. Curvature thereby becomes a purely geometric control knob for collective spontaneous emission with a sign sensitivity, one that photon-mediated interactions in a dielectric lack as their potential is even in the curvature~\cite{carmi2025photon}.

Since the geometric potentials weight the spatial profile of each mode differently, combining phase matching with a designed curvature landscape may allow individual super- and subradiant channels to be targeted, which invites a study of collective-light--collective-matter interactions~\cite{jalali2024collective, jalali2026collective} and polariton-assisted superradiance~\cite{chuang2026polaron} on curved surfaces. A second direction is the anti-Hermitian part of Eq.~\eqref{eq:lossy_equation}, through which a designed curvature landscape controls where and along which direction an SPP is damped, giving access to non-Hermitian phenomena~\cite{ashida2020non, Bergholtz2021}. Beyond plasmonics, our results suggest that the order at which a geometric potential enters depends on the symmetry of the transverse mode profile, which should extend to other asymmetrically confined waves such as surface phonon polaritons, and to acoustic or quantum-mechanical surface modes. At the same time, the position-, direction-, and sign-dependent dispersion imprinted by $H$ and $\sigma^{ab}$ turns a curved metal into an effective geometry for the SPP, of possible use for analog gravity~\cite{xu2026photon, zhang2026curved} or curved-space topological phases~\cite{lustig2017curved}.

\textit{Acknowledgments.}---We thank Federico Roccati for related discussions. We acknowledge exchange within the Max Planck – uOttawa Centre for
Extreme and Quantum Photonics.
We further acknowledge funding from the Max Planck Society's Lise Meitner Excellence Program 2.0, and acknowledge acceptance of an early-stage version of this work at META 2026 in Dublin.

\textit{Data Availability.}---The data and codes of the numerics are available upon request.

\nocite{misner1973gravitation, deserno2004notes, lai2018electromagnetic, sommerfeld1949partial, berenger1994perfectly}

\bibliography{references}

\onecolumngrid
\section*{End Matter}
\twocolumngrid
\balance

\textit{Outline of the main derivation.}---To extend the SPP physics to curved interfaces, we introduce coordinates adapted to the interface, in which the three-dimensional metric reads $g_{ab}=\gamma_{ab}-2\eta h_{ab}$ to first order in $\eta$~\cite{da1981quantum}, with $\gamma_{ab}$ the intrinsic surface metric. Since the gauge freedom allows the tangential components of the four-potential to vanish on a flat interface~\cite{maier2007plasmonics}, see the SM~\cite{supp}, any tangential field appearing below is generated by the curvature. The mode extends over $\kappa_{d,m}^{-1}$ into the two media and therefore samples the curvature at a finite distance from the interface, at most as $\sim H/\kappa_d$ on the dielectric side, which is $\alpha$ up to the material factor $\sqrt{-\epsilon_m/\epsilon_d}$.

Expanding Maxwell's equations in the Lorenz gauge, the wave equation for the normal field component, rescaled as $\mathcal{A}\equiv\sqrt{\Omega}A$ with $\Omega=1-2H\eta$~\cite{batz2008linear}, acquires an extra term $2\eta D_h$, with the operator $D_h\equiv h^{bc}\nabla_b\nabla_c=H\Delta_\gamma+\sigma^{bc}\nabla_b\nabla_c$ contracting covariant derivatives along the surface. Curvature further generates tangential components of the vector field and modifies the boundary matching conditions at $\eta=0$ to
\begin{equation}\label{eq:em_boundary}
\left.\frac{\partial_\eta\mathcal{A}^\eta_m}{\epsilon_m}-\frac{\partial_\eta\mathcal{A}^\eta_d}{\epsilon_d}\right|_{\eta=0}=\frac{\epsilon_d-\epsilon_m}{\epsilon_m\epsilon_d}\left.\left(H\mathcal{A}^\eta-\nabla_a\mathcal{A}^a\right)\right|_{\eta=0},
\end{equation}
where the right-hand side vanishes on a flat interface with $H=0$ and $\nabla_a\mathcal{A}^a=0$. Together with $2\eta D_h$, which the thin-layer limit discards, Eq.~\eqref{eq:em_boundary} is the mathematical origin of the first-order-in-curvature potentials.

We then solve order by order: at $\mathcal{O}(\alpha)$ the normal profile acquires a polynomial-in-$\eta$ correction to its exponential, whereas in locally adapted coordinates, e.g., in a Monge parametrization, the intrinsic-geometry operators, the Laplace-Beltrami operator $\Delta_\gamma$, and the Christoffel symbols depart from their flat counterparts at $\mathcal{O}(\alpha^2)$, and hence do not contribute at the order we work in. Matching the fields at each order to the boundary conditions, including the curvature-induced tangential divergence $\nabla_a\mathcal{A}^a$, then yields Eq.~\eqref{eq:main_equation}, with all technical steps given in the SM~\cite{supp}.

\textit{Recovery of known results for spheres.}---A convex metal sphere of radius $R$ has principal curvatures $\kappa_{1,2}=-1/R$, yielding $H=-1/R$ and $\sigma^{ab}=0$. Thus Eq.~\eqref{eq:main_equation} reduces to $[\Delta_\gamma+k_{\rm eff}^2]\psi=0$ with
\begin{equation}\label{eq:em_k_eff_sphere}
k_{\rm eff}\approx k_{\rm spp}\left(1-\frac{ \epsilon_d^2 + \epsilon_d \epsilon_m + \epsilon_m^2}{2k_0R\,\epsilon_m\epsilon_d\sqrt{-(\epsilon_d + \epsilon_m)}}\right).
\end{equation}
This blue-shifted correction matches the ray-asymptotic dispersion of Ref.~\cite{perel2011asymptotics}. It also recovers Ref.~\cite{ancey2009surface} in the symmetric-decay limit ($\epsilon_m\to-\epsilon_d$), giving $k_{\rm eff}\approx k_{\rm spp}(1+1/[2k_0R\sqrt{-(\epsilon_d+\epsilon_m)}])$. 

\textit{Recovery of known results for cylinders.}---A cylinder (axisymmetric around $z$ and with arclength $s$) of radius $R$ has principal curvatures $\kappa_z=0$, $\kappa_s=-1/R$, a flat intrinsic metric, $H=-1/(2R)$, and $\sigma^{zz}=-\sigma^{ss}=1/(2R)$. Equation~\eqref{eq:main_equation} reads
\begin{equation}\label{eq:em_cylinder_full}
\left(1 + \frac{C_\sigma}{2R}\right)\partial_z^2\psi + \left(1 - \frac{C_\sigma}{2R}\right)\partial_s^2\psi + \left(k_{\rm spp}^2 - \frac{C_H}{2R}\right)\psi = 0.
\end{equation}
Under a slowly varying envelope $\psi=F(z,s)e^{-ik_{\rm spp}z}$, and assuming the envelope also varies slowly along $s$, this reduces to the paraxial Eq.~(9) of Ref.~\cite{della2010geometric}
\begin{equation}\label{eq:em_paraxial_result}
-i\bar\lambda_0\,\partial_z F = -\frac{\bar\lambda_0^2}{2n_e}\,\partial_s^2 F - \frac{\bar\lambda_0\, n_e}{2R\sqrt{-(\epsilon_m + \epsilon_d)}}\,F,
\end{equation}
with $\bar\lambda_0=1/k_0$ and $n_e=k_{\rm spp}/k_0$. Alternatively, substituting $\psi=e^{ik_zz+im\theta}$ (with $k_\theta=m/R$) yields the momentum ellipse for convex (upper signs) and concave (lower signs) cylinders
\begin{equation}\label{eq:em_momentum_ellipse}
k_z^2 \left( 1 \pm \frac{C_\sigma}{2R} \right) + k_\theta^2 \left( 1 \mp \frac{C_\sigma}{2R} \right) = k_{\text{spp}}^2 \mp \frac{C_H}{2R},
\end{equation}
which degenerates to a circle when $C_\sigma=0$. Expanding to leading order in $1/R$, and defining $V\equiv k_0R\sqrt{-(\epsilon_d+\epsilon_m)}$, yields 
\begin{equation}\label{eq:em_cyl_limits_k_z}
    k^2_z(m)=k^2_{\rm spp}\left(1\pm\frac{1}{V}\right)-\frac{m^2}{R^2},
\end{equation}
and separately setting $k_z=0$ (pure azimuthal propagation) gives
\begin{equation}\label{eq:em_cyl_limits_k_theta}
    k_\theta = k_{\text{spp}} \sqrt{ 1 \mp \frac{(\epsilon_d + \epsilon_m)^2}{\epsilon_d \epsilon_m V} },
\end{equation}
in agreement with Ref.~\cite{spittel2015curvature}.

\onecolumngrid
\clearpage

\onecolumngrid

\setcounter{secnumdepth}{4}
\setcounter{section}{0}
\setcounter{figure}{0}
\setcounter{equation}{0}
\renewcommand\thesection{S\arabic{section}}
\renewcommand\thefigure{S\arabic{figure}}
\renewcommand\theequation{S\arabic{equation}}
\setlength{\belowcaptionskip}{0pt}

\begin{center}
  \textbf{\large Supplemental Material:\\[0.2cm]
  Birefringent Curvature Control of Surface Plasmons and Collective Emission
  }\\[.2cm]
  Florian B\"onsel$^{*}$ and Flore K.~Kunst$^{\dagger}$\\[.1cm]
  {\itshape $^1$Max Planck Institute for the Science of Light, 91058 Erlangen, Germany\\
  $^2$Department of Physics, Friedrich-Alexander-Universit\"at Erlangen-N\"urnberg, 91058 Erlangen, Germany\\}
  (Dated: \today)
\end{center}
\vspace{0.5cm}

\makeatletter
\def\l@subsection#1#2{}
\def\l@subsubsection#1#2{}
\def\l@paragraph#1#2{}
\makeatother

This Supplemental Material contains detailed derivations, as well as reviews and recapitulations of established physics, which are re-derived here for a full picture supporting reproducibility.

\SMTOCoff
\tableofcontents
\addtocontents{toc}{\protect\SMTOCon}

\section{Flat-interface SPP mode and gauge conditions}\label{sec:flat_spp_and_gauge}

Here, we give the flat-interface solution underlying Eq.~(1) of the main text, and motivate the gauge choice used in sections~\ref{supp:wave_equation_curved_space}--\ref{supp:derivation_tangential_divergence}. Here, we take non-magnetic materials, real $\epsilon_m$, and real $\epsilon_d$. Denoting by $\eta$ the coordinate normal to the interface, the permittivity profile reads
\begin{equation} \label{eq:permitivity_profile}
    \epsilon(\eta) = \begin{cases} \epsilon_d & \text{for } \eta > 0 \quad (\text{dielectric}), \\ \epsilon_m & \text{for } \eta < 0 \quad (\text{metal}), \end{cases}
\end{equation}
with $\mathrm{Re}\left[\epsilon_m\right] < 0 < \epsilon_d$ required for the mode to exist. The inverse localization lengths of Eq.~(1) of the main text read in closed form
\begin{equation}\label{eq:kappa_wavenumber}
    \kappa_{d,m} = \sqrt{k_{\rm spp}^2 - \epsilon_{d,m} k_0^2} = \frac{k_0\,|\epsilon_{d,m}|}{\sqrt{-(\epsilon_m+\epsilon_d)}}.
\end{equation}

The physical fields of the transverse magnetic (TM) mode are given as Eqs.~(2.10) and~(2.11) in Ref.~\cite{maier2007plasmonics}. One four-vector potential $\mathbf{\mathcal{A}} = \left(\phi,\mathcal{A}^1, \mathcal{A}^2, \mathcal{A}^\eta \right) = \left(\phi, 0, 0, \mathcal{A}^z \right)$ reproducing them is
\begin{equation}\label{eq:A_z_flat}
    \mathcal{A}^z = \frac{i\phi_0 \mu_0}{k_{\rm spp}}e^{ik_{\rm spp}x} \begin{cases} e^{-\kappa_d z}, & z > 0,\\ e^{\kappa_m z}, & z < 0, \end{cases}
\end{equation}
and
\begin{equation}\label{eq:phi_flat}
    \phi = \frac{\phi_0}{\omega \epsilon_0 k_{\rm spp}}e^{ik_{\rm spp}x}
    \begin{cases} 
    -\frac{\kappa_d}{\epsilon_d}e^{-\kappa_d z}, & z > 0,\\ \frac{\kappa_m}{\epsilon_m}e^{\kappa_m z}, & z < 0, 
    \end{cases}
\end{equation}
with $\epsilon_0$ the vacuum electric permittivity and $\mu_0$ the vacuum magnetic permeability. To compare with Ref.~\cite{maier2007plasmonics}, we use the notation $A_1 = A_2 = \phi_0$ to avoid confusion with the vector field components in our work. Both, Eq.~\eqref{eq:A_z_flat} and Eq.~\eqref{eq:phi_flat}, are continuous at $z=0$, and they satisfy the Lorenz gauge,
\begin{equation}\label{eq:lorenz_gauge_flat}
    \phi(z) = \frac{\left(\nabla \cdot \mathbf{\mathcal{A}}\right)(z)}{i\omega\mu_0\epsilon_0\epsilon(z)} = \frac{\left( \partial_z \mathcal{A}^z\right)(z)}{i\omega\mu_0\epsilon_0\epsilon(z)}.
\end{equation}
The normal component is separable,
\begin{equation}\label{eq:separation_Az}
    \mathcal{A}^z_i = f_i(z)\,\psi(x,y), \qquad f_i(z) = {\rm exp}(-\kappa_i|z|),
\end{equation}
with $i=m$ ($i=d$) in the metal (dielectric), which is why the Green's function in  Section~\ref{supp:derivation_constant_C0} factorizes. The tangential divergence $\nabla_a\mathcal{A}^a = \partial_x\mathcal{A}^x + \partial_y\mathcal{A}^y$ vanishes since $\mathcal{A}^x = \mathcal{A}^y = 0$, which is why the right-hand side (RHS) of Eq.~(2) of the main text vanishes on a flat interface. The tangential components generated by curvature, and with them a transverse electric field that does not exist for flat surfaces~\cite{maier2007plasmonics}, are computed to first order in Section~\ref{supp:derivation_tangential_divergence}.

\section{Electromagnetic wave equation in curved space}\label{supp:wave_equation_curved_space}
To arrive at the final equation Eq.~(3) presented in the main text, we start with the covariant formulation of the vector potential wave equation in the Lorenz gauge using SI-units from Ref.~\cite{misner1973gravitation},  Eq.~22.19(d), which reads
\begin{equation}\label{eq:wave_equation_mtw}
    -\nabla_{\beta}\nabla^{\beta}A^{\alpha}+{R^{\alpha}}_{\beta}A^{\beta}=\mu_0  J^{\alpha},
\end{equation}
with $\nabla_{\beta}$ the covariant derivative, $A^{\alpha}$ the vector potential, ${R^{\alpha}}_{\beta}$ the Ricci tensor, $\mu_0$ the permeability of the vacuum, $J^{\alpha}$ the current density, and
using the space-time index $\alpha,\beta \in\{0,1,2,3\}$ and the metric signature $(-,+,+,+)$. Now we have to make clear that for the case of a curved interface, the interface itself is an embedded surface $\mathcal{S}$ acting as a constraining geometry for the SPP mode within the ambient 3D space. While $\mathcal{S}$ is curved, the 3D ambient space is flat, so $R^i_m=0$ with Roman letters referring to the spatial indices $i\in \{1,2,3\}$. Following the standard approach laid out in Ref.~\cite{da1981quantum}, and introducing a coordinate system describing $\mathcal{S}$ with two tangential coordinates and a normal coordinate will reveal the impact of the curved constraining geometry on the SPP physics.

The Christoffel symbols are defined as
\begin{equation}
    \Gamma^{\alpha}_{\mu\nu}=\frac{1}{2}g^{\alpha\beta}\left(\partial_\mu g_{\beta\nu}+\partial_\nu g_{\mu\beta} - \partial_\beta g_{\mu\nu}\right)
\end{equation}
within a coordinate-basis framework with $g_{\alpha \beta}$ the metric tensor. We assume a static metric, i.e.,
\begin{equation}
    g_{0i}=0,\qquad \partial_0g_{ik}=0,\qquad \partial_ig_{00}=0.
\end{equation} 
This means that all Christoffel symbols, except the purely spatial ones, vanish. Due to the metric-compatibility ($\nabla_i g^{jk}=0$)~\cite{deserno2004notes}, we can write
\begin{equation}
    -\nabla_{\beta}\nabla^{\beta}A^{\alpha}=-g^{\beta\mu}\nabla_{\beta}\nabla_{\mu}A^{\alpha}.
\end{equation}
Expanding this yields
\begin{equation}
    \begin{aligned}
    -g^{\beta\mu}\nabla_{\beta}\nabla_{\mu}A^{\alpha}=
    &-g^{\beta\mu}\partial_{\beta}\partial_{\mu}A^{\alpha}-g^{\beta\mu}(\partial_{\beta}{\Gamma^{\alpha}}_{\sigma\mu})A^{\sigma}-g^{\beta\mu}{\Gamma^{\alpha}}_{\sigma\mu}\partial_{\beta}A^{\sigma}-g^{\beta\mu}{\Gamma^{\alpha}}_{\rho\beta}\partial_{\mu}A^{\rho}\\
    &-g^{\beta\mu}{\Gamma^{\alpha}}_{\rho\beta}{\Gamma^{\rho}}_{\sigma\mu}A^{\sigma}+g^{\beta\mu}{\Gamma^{\lambda}}_{\mu\beta}\partial_{\lambda}A^{\alpha}+g^{\beta\mu}{\Gamma^{\lambda}}_{\mu\beta}{\Gamma^{\alpha}}_{\sigma\lambda}A^{\sigma}.
\end{aligned}
\end{equation}
Splitting into temporal and spatial parts, we get
\begin{equation}
    \begin{aligned}\label{eq:vector_laplacian_expanded}
    -g^{\beta\mu}\nabla_{\beta}\nabla_{\mu}A^{\alpha}=
    &-g^{00}\partial_0^2 A^\alpha 
    -g^{jk}\partial_j\partial_k A^\alpha 
    - g^{jk}(\partial_j \Gamma^\alpha_{km})A^m 
    - g^{jk}\Gamma^\alpha_{km}\partial_j A^m
    -g^{jk}\Gamma^\alpha_{j m}\partial_k A^m\\
    &- g^{jk}\Gamma^\alpha_{jl}\Gamma^l_{k m}A^m 
    + g^{jk}\Gamma^m_{jk}\partial_m A^\alpha 
    + g^{jk}\Gamma^m_{jk}\Gamma^\alpha_{ml}A^l,
\end{aligned}
\end{equation}
as every Christoffel symbol with an index $0$ vanishes. In the following, we look at $\alpha=i$. For a monochromatic ansatz $A^\alpha(t, \mathbf{r}) = \tilde{A}^\alpha(\mathbf{r}) e^{-i\omega t}$ and the refractive index of the medium $n_0 = \frac{1}{\sqrt{-g_{00}}}$, the first term on the RHS of Eq.~\eqref{eq:vector_laplacian_expanded} becomes 
\begin{equation}
    -g^{00}\partial_0^2 A^i=g^{00}\omega^2  A^i=-n^2_0\omega^2 A^i=-k^2_0n^2_0 A^i.
\end{equation}
The second and seventh term on the RHS of Eq.~\eqref{eq:vector_laplacian_expanded} are the negative of the scalar Laplacian
\begin{equation}
    -g^{jk}\partial_j\partial_k A^i+g^{jk}\Gamma^m_{jk}\partial_m A^i=- \frac{1}{\sqrt{g}}\partial_j\!\left(\sqrt{g}\,g^{jk}\partial_k A^i\right)=-\Delta A^i,
\end{equation}
and the third, fourth, and eighth term on the RHS of Eq.~\eqref{eq:vector_laplacian_expanded} can be summed  as
\begin{equation}
    -g^{jk}(\partial_j \Gamma^i_{km})A^m -g^{jk}\Gamma^i_{km}\partial_j A^m+ g^{jk}\Gamma^m_{jk}\Gamma^i_{ml}A^l=-\frac{1}{\sqrt{g}}\partial_j\left(\sqrt{g}g^{jk}\Gamma^i_{kl}A^l \right).
\end{equation}
Assembling all terms, multiplying with $-1$, we can write the wave equation as
\begin{equation}\label{eq:wave_equation_general}
    \mathcal{L}^i = -\mu_0 J^i,
\end{equation}
with
\begin{equation} \label{eq:app_wave_eq}
    \mathcal{L}^i \equiv \underbrace{\left(\Delta + k^2_0\epsilon\right)A^i}_{\rm Term\,I} 
    +\underbrace{\frac{1}{\sqrt{g}}\partial_j\left(\sqrt{g}g^{jk}\Gamma^i_{kl}A^l \right)
    +g^{jk}\Gamma^i_{j m}\partial_k A^m
    + g^{jk}\Gamma^i_{jl}\Gamma^l_{k m}A^m}_{\rm Term\,II},
\end{equation}
and $n_0^2=\epsilon$.

Applying now the parametrization of the surface $\mathcal{S}$, we follow Ref.~\cite{da1981quantum} and introduce $\mathbf{R}(q^1, q^2, \eta) = \mathbf{r}(q^1, q^2) + \eta\, \hat{\mathbf{n}}(q^1, q^2)$. Then, defining the intrinsic metric $\gamma_{ab}$ and the second fundamental form $h_{ab}$, the full metric can be written as 
\begin{equation}
    g_{ab} = \gamma_{ab} - 2 \eta h_{ab},\qquad g_{\eta\eta}=1,\qquad g_{a\eta}=0,
\end{equation}
since we truncate all terms at linear order in curvature, with $a\in\{1,2\}$ the tangential indices. The inverse metric reads $g^{ab} \approx \gamma^{ab} + 2 \eta h^{ab}$, and the volume element is $\sqrt{g} = \sqrt{\gamma} \Omega$ with $\Omega \approx 1 - 2 \eta H$. 
Throughout this work, we keep $\eta$, while dropping any terms quadratic in the curvature (such as $H^2$, $H h_{ab}$, or $h_{ac}h^c_b$) as well as the spatial derivatives of the curvature ($\partial_b H \to 0$, $\partial_b h_{cd} \to 0$), cf.~Ref.~\cite{batz2008linear}. Operators or symbols with only indices $a,b,c,d,...$ are defined intrinsically on the surface.  We now calculate the Christoffel symbols required for the wave equation using $\partial_\eta g_{ab} = -2 h_{ab}$. They are calculated to
\begin{equation}
    \begin{aligned}
    &\Gamma^\eta_{ab} = h_{ab},\\
    &\Gamma^a_{\eta b}=\Gamma^a_{b\eta}=-h^a_b,\\
    &\Gamma^a_{bc}=\tilde\Gamma^a_{bc}+2\eta  h^{ad}\gamma_{de}\Gamma^e_{bc}\approx \tilde\Gamma^a_{bc},
\end{aligned}
\end{equation}
with the two-dimensional surface Christoffel symbols $\tilde \Gamma^a_{bc} \equiv  \frac{1}{2} \gamma^{ad} (\partial_b \gamma_{dc} + \partial_c \gamma_{db} - \partial_d \gamma_{bc})$. In the following, however, we just write $\Gamma^a_{bc}$. All other symbols ($\Gamma^\eta_{\eta\eta}$, $\Gamma^a_{\eta\eta}$, $\Gamma^\eta_{\eta\eta}$) are zero. In the subsequent derivations, we make use of the Mainardi-Codazzi equations $\nabla_b h_{ac} = \nabla_c h_{ab}$~\cite{deserno2004notes} several times to drop higher-order terms, such as $\nabla_a h^{ab}$, as
\begin{equation} \label{eq:app_mainardi_codazzi}
    \begin{aligned}
    \nabla_b h_{ac} &= \nabla_c h_{ab},\\
    \gamma^{ac} \nabla_b h_{ac} &= \gamma^{ac} \nabla_c h_{ab},\\
    \nabla_b (\gamma^{ac} h_{ac}) &= \nabla_c (\gamma^{ac} h_{ab}),\\
    \nabla_b ( h^a_{a}) &= \nabla_c (h^c_{b}),\\
    \nabla_b (2H) &= \nabla_c (h^c_{b}),\\
    0\approx 2\partial_b H &= \nabla_c (h^c_{b}).
\end{aligned}
\end{equation}
In line with the standard approaches for the physics on curved surfaces~\cite{da1981quantum}, we use the rescaling of the field $A^i=\mathcal{A}^i/\sqrt{\Omega}$. In the next sections, we will need the following
\begin{equation}
    \partial_k A^m 
    = \partial_k \left( \frac{\mathcal{A}^m}{\sqrt{\Omega}} \right) 
    = 
    \frac{1}{\sqrt{\Omega}} \partial_k \mathcal{A}^m 
    -\frac{1}{2}\frac{\partial_k \Omega}{\Omega^{3/2}} \mathcal{A}^m.
\end{equation} 
Using $\partial_\eta \Omega = -2H$ and $\partial_c  \Omega  \approx 0$, we approximate this to
\begin{equation}
    \partial_c A^m \approx \frac{1}{\sqrt{\Omega}} \partial_c \mathcal{A}^m, \qquad \partial_\eta A^m \approx \frac{1}{\sqrt{\Omega}} \partial_\eta \mathcal{A}^m + H \mathcal{A}^m.
\end{equation}

\subsection{Term I}\label{supp:term_I}
We start with the Laplacian in term I in Eq.~\eqref{eq:app_wave_eq}. We can split it into tangential and normal contributions as 
\begin{equation} 
\Delta A^i = \underbrace{\frac{1}{\sqrt{\gamma}}\partial_a\!\left(\sqrt{\gamma}\,g^{ab}\partial_b A^i\right)}_{\displaystyle\Delta_{\parallel} A^i} + \underbrace{\frac{1}{\Omega}\partial_\eta\!\left(\Omega\,\partial_\eta A^i\right)}_{\displaystyle\Delta_{\perp} A^i}, 
\end{equation} 
for which we already used that $\partial_a\Omega \approx0$ and that $\gamma$ does not depend on $\eta$. We now expand the inverse metric $g^{ab} = \gamma^{ab} + 2h^{ab}\eta + \mathcal{O}(\eta^2)$. The tangential part is
\begin{equation}
    \Delta_{\parallel} A^i=\frac{1}{\sqrt{\gamma}}\partial_a\!\left(\sqrt{\gamma}\left(\gamma^{ab} +2\eta h^{ab}\right)\partial_b A^i\right)\approx \Delta_\gamma A^i+2\eta\,\frac{1}{\sqrt{\gamma}}\partial_a\!\left(\sqrt{\gamma}\,h^{ab}\partial_b A^i\right)
\end{equation}
with $\Delta_\gamma = \frac{1}{\sqrt{\gamma}}\partial_a(\sqrt{\gamma}\,\gamma^{ab}\partial_b)$ as the Laplace-Beltrami operator based on the intrinsic surface metric. With the definition of the intrinsic surface divergence
\begin{equation}
    \bar \nabla_a V^ a\equiv \frac{1}{\sqrt{\gamma}} \partial_a \left( \sqrt{\gamma} \, V^a \right),
\end{equation}
and identifying $V^a=h^{ab}\partial_b A^i$, since $A^i$ is just a scalar on the 2D surface, we get
\begin{equation}
    \frac{1}{\sqrt{\gamma}}\partial_a\!\left(\sqrt{\gamma}\,h^{ab}\partial_b A^i\right)
    = \nabla_a\left(h^{ab}\nabla_b \mathcal{A}^i \right)
    =\left(\nabla_a h^{ab} \right)\nabla_b A^i + h^{ab}\nabla_a\nabla_b \mathcal{A}^i\approx h^{ab}\nabla_a\nabla_b A^i,
\end{equation}
where we dropped $\nabla_a h^{ab}$ using the Codazzi-Mainardi equations, cf. Eq.~\eqref{eq:app_mainardi_codazzi}. The tangential Laplacian, therefore, reads
\begin{equation} 
\sqrt{\Omega}\,\Delta_{\parallel} A^i = \left[\Delta_\gamma + 2\eta\,h^{ab}\nabla_a\nabla_b\right]\mathcal{A}^i + \mathcal{O}(H^2). \label{eq:Delta_parallel} 
\end{equation}
For the normal part, we get
\begin{equation}\label{eq:Delta_perp_correct} 
\sqrt{\Omega}\,\Delta_\perp A^i =\partial_\eta^2\mathcal{A}^i + \mathcal{O}(H^2). 
\end{equation} 
Adding Eq.~\eqref{eq:Delta_perp_correct} with the tangential result of Eq.~\eqref{eq:Delta_parallel} yields 
\begin{equation}\label{eq:TermI_general}
\sqrt{\Omega}\!\left[\Delta + k_0^2\epsilon(\eta)\right]\!A^i \approx  \left[\Delta_\gamma + \partial_\eta^2 + k_0^2\epsilon\right]\!\mathcal{A}^i + 2\eta\,h^{ab}\nabla_a\nabla_b\mathcal{A}^i.
\end{equation}

\subsection{Term II}\label{supp:term_II}
With the expansion of the metric $g$, the first term of term II in Eq.~\eqref{eq:app_wave_eq} can be written as
\begin{equation}
    \begin{aligned}
    T_1^i &= \frac{1}{\Omega} \partial_\eta \left( \sqrt{\Omega} \Gamma^i_{\eta l} \mathcal{A}^l \right) 
    + \frac{1}{\sqrt{\gamma}\sqrt{\Omega}} \partial_b \left( \sqrt{\gamma} g^{bc} \Gamma^i_{cl} \mathcal{A}^l \right)\\
    & = \frac{1}{\Omega} \partial_\eta \left[ \sqrt{\Omega} \left( \Gamma^i_{\eta d}\mathcal{A}^d + \Gamma^i_{\eta\eta}\mathcal{A}^\eta \right) \right] + \frac{1}{\sqrt{\gamma}\sqrt{\Omega}} \partial_b \left[ \sqrt{\gamma} g^{bc} \left( \Gamma^i_{cd}\mathcal{A}^d + \Gamma^i_{c\eta}\mathcal{A}^\eta \right) \right]\\
    &= \frac{1}{\Omega} \partial_\eta \left( \sqrt{\Omega} \Gamma^i_{\eta d}\mathcal{A}^d \right) + \frac{1}{\sqrt{\gamma}\sqrt{\Omega}} \partial_b \left[ \sqrt{\gamma} g^{bc} \left( \Gamma^i_{cd}\mathcal{A}^d + \Gamma^i_{\eta c}\mathcal{A}^\eta \right) \right].
\end{aligned}
\end{equation}
Therefore,
\begin{equation}
    \sqrt{\Omega}\,T_1^i=\frac{1}{\sqrt{\Omega}} \partial_\eta \left( \sqrt{\Omega} \Gamma^i_{\eta d}\mathcal{A}^d \right) + \frac{1}{\sqrt{\gamma}} \partial_b \left[ \sqrt{\gamma} g^{bc} \left( \Gamma^i_{cd}\mathcal{A}^d + \Gamma^i_{\eta c}\mathcal{A}^\eta \right) \right].
\end{equation}
Now for $i=a$, we get
\begin{equation}
    \begin{aligned}
    \sqrt{\Omega}\,T_1^a&= \frac{1}{\sqrt{\Omega}} \partial_\eta \left( \sqrt{\Omega} (-h^a_d) \mathcal{A}^d \right) + \frac{1}{\sqrt{\gamma}} \partial_b \left[ \sqrt{\gamma} g^{bc} \left( \Gamma^a_{cd}\mathcal{A}^d - h^a_c \mathcal{A}^\eta \right) \right]\\
    &= -\frac{h^a_d}{\sqrt{\Omega}} \partial_\eta \left( \sqrt{\Omega} \mathcal{A}^d \right) + \frac{1}{\sqrt{\gamma}} \partial_b \left[ \sqrt{\gamma} g^{bc} \left( \Gamma^a_{cd}\mathcal{A}^d - h^a_c \mathcal{A}^\eta \right) \right]\\
    &\approx -\frac{h^a_d}{\sqrt{\Omega}} \partial_\eta \left( (1 - \eta H) \mathcal{A}^d \right) +  \frac{1}{\sqrt{\gamma}} \partial_b \left[ \sqrt{\gamma} g^{bc} \left( \Gamma^a_{cd}\mathcal{A}^d - h^a_c \mathcal{A}^\eta \right) \right]\\
    &= \frac{1}{\sqrt{\Omega}} \left( h^a_d H \mathcal{A}^d - h^a_d \partial_\eta \mathcal{A}^d + \eta h^a_d H \partial_\eta \mathcal{A}^d \right)+ \frac{1}{\sqrt{\gamma}} \partial_b \left[ \sqrt{\gamma} g^{bc} \left( \Gamma^a_{cd}\mathcal{A}^d - h^a_c \mathcal{A}^\eta \right) \right]\\
    &\approx \frac{1}{\sqrt{\Omega}} \left( - h^a_d \partial_\eta \mathcal{A}^d \right)+ \frac{1}{\sqrt{\gamma}} \partial_b \left[ \sqrt{\gamma} g^{bc} \left( \Gamma^a_{cd}\mathcal{A}^d - h^a_c \mathcal{A}^\eta \right) \right]\\
    &\approx- h^a_d \partial_\eta \mathcal{A}^d+ \frac{1}{\sqrt{\gamma}} \partial_b \left[ \sqrt{\gamma} g^{bc} \left( \Gamma^a_{cd}\mathcal{A}^d - h^a_c \mathcal{A}^\eta \right) \right]\\
    &= - h^a_d \partial_\eta \mathcal{A}^d + \frac{1}{\sqrt{\gamma}} \partial_b \left[ \sqrt{\gamma} \left(\gamma^{bc}+2\eta h^{bc} \right)\left( \Gamma^a_{cd}\mathcal{A}^d - h^a_c \mathcal{A}^\eta \right) \right]\\
    &= - h^a_d \partial_\eta \mathcal{A}^d+ \frac{1}{\sqrt{\gamma}} \partial_b \left[ \sqrt{\gamma} \left( \gamma^{bc} \Gamma^a_{cd}\mathcal{A}^d - h^{ab} \mathcal{A}^\eta + 2\eta h^{bc} \Gamma^a_{cd}\mathcal{A}^d \right) \right].
\end{aligned}
\end{equation}
For the normal $i=\eta$ component, we have
\begin{equation}
    \begin{aligned}
    \sqrt{\Omega}\,T_1^\eta &= \frac{1}{\sqrt{\Omega}} \partial_\eta \left( \sqrt{\Omega} (0) \mathcal{A}^d \right) + \frac{1}{\sqrt{\gamma}} \partial_b \left[ \sqrt{\gamma} g^{bc} \left( h_{cd}\mathcal{A}^d + (0)\mathcal{A}^\eta \right) \right]\\
    &= \frac{1}{\sqrt{\gamma}} \partial_b \left( \sqrt{\gamma} g^{bc} h_{cd}\mathcal{A}^d \right)\\
    &\approx \frac{1}{\sqrt{\gamma}} \partial_b \left( \sqrt{\gamma} \gamma^{bc} h_{cd}\mathcal{A}^d \right)\\
    &= \frac{1}{\sqrt{\gamma}} \partial_b \left( \sqrt{\gamma} h^b_d \mathcal{A}^d \right)\\
    & = (\nabla_b h^b_d) \mathcal{A}^d + h^b_d (\nabla_b \mathcal{A}^d)\\
    &\approx h^b_d (\nabla_b \mathcal{A}^d).
\end{aligned}
\end{equation}
We now continue with the second term of term II in Eq.~\eqref{eq:app_wave_eq}:
\begin{equation}
    \begin{aligned}
    T_2^i 
    &= g^{jk}\Gamma^i_{j m}\partial_k A^m\\
    &=\Gamma^i_{\eta m}\partial_\eta A^m + g^{bc}\Gamma^i_{bm}\partial_c A^m\\
    &= \left( \Gamma^i_{\eta d}\partial_\eta A^d + \Gamma^i_{\eta\eta}\partial_\eta A^\eta \right) + g^{bc} \left( \Gamma^i_{bd}\partial_c A^d + \Gamma^i_{b\eta}\partial_c A^\eta \right)\\
    &= \Gamma^i_{\eta d}\partial_\eta A^d + g^{bc} \left( \Gamma^i_{bd}\partial_c A^d + \Gamma^i_{b\eta}\partial_c A^\eta \right).
\end{aligned}
\end{equation}
We set $i=\eta$ and substitute the Christoffel symbols $\Gamma^\eta_{\eta d} = 0$, $\Gamma^\eta_{bd} = h_{bd}$, and $\Gamma^\eta_{b\eta} = 0$, so
\begin{equation}
    \begin{aligned}
    \sqrt{\Omega}\,T_2^\eta &
    = \sqrt{\Omega}\,g^{bc} h_{bd} \partial_c A^d\\
    &=\sqrt{\Omega}\, g^{bc} h_{bd} \left( \frac{1}{\sqrt{\Omega}} \partial_c \mathcal{A}^d \right)\\
    &=g^{bc} h_{bd} \left(\partial_c \mathcal{A}^d \right)\\
    &\approx  h^c_{d} \partial_c \mathcal{A}^d.
\end{aligned}
\end{equation}
For $i=a$, we get
\begin{equation}
    \begin{aligned}
   \sqrt{\Omega}\, T_2^a 
    &= -\sqrt{\Omega}\,h^a_d \partial_\eta A^d + \sqrt{\Omega}\,g^{bc} \left( \Gamma^a_{bd}\partial_c A^d - h^a_b \partial_c A^\eta \right)\\
    & = -\sqrt{\Omega}\,h^a_d \left( \frac{1}{\sqrt{\Omega}} \partial_\eta \mathcal{A}^d + H \mathcal{A}^d \right) + \sqrt{\Omega}\,g^{bc} \Gamma^a_{bd} \left( \frac{1}{\sqrt{\Omega}} \partial_c \mathcal{A}^d \right) - \sqrt{\Omega}\,g^{bc} h^a_b \left( \frac{1}{\sqrt{\Omega}} \partial_c \mathcal{A}^\eta \right)\\
    & \approx  -h^a_d \partial_\eta \mathcal{A}^d + g^{bc} \Gamma^a_{bd} \left( \partial_c \mathcal{A}^d \right) - g^{bc} h^a_b \left(\partial_c \mathcal{A}^\eta \right)\\
    & \approx  -h^a_d \partial_\eta \mathcal{A}^d + \gamma^{bc} \Gamma^a_{bd} \partial_c \mathcal{A}^d + 2\eta h^{bc} \Gamma^a_{bd} \partial_c \mathcal{A}^d - g^{bc} h^a_b \left( \partial_c \mathcal{A}^\eta \right)\\
    & \approx  -h^a_d \partial_\eta \mathcal{A}^d +  \gamma^{bc} \Gamma^a_{bd} \partial_c \mathcal{A}^d + 2\eta h^{bc} \Gamma^a_{bd} \partial_c \mathcal{A}^d - h^{ac} \partial_c \mathcal{A}^\eta.
\end{aligned}
\end{equation}
The third and last term of term II in Eq.~\eqref{eq:app_wave_eq} is
\begin{equation}
    \begin{aligned}
    \sqrt{\Omega}\,T_3^i 
    & = g^{jk}\Gamma^i_{jl}\Gamma^l_{k m}\mathcal{A}^m\\
    &= \left( \Gamma^i_{\eta l}\Gamma^l_{\eta m} + g^{bc}\Gamma^i_{bl}\Gamma^l_{cm} \right) \mathcal{A}^m\\
    &= \left[ \left( \Gamma^i_{\eta d}\Gamma^d_{\eta m} + \Gamma^i_{\eta \eta }\Gamma^\eta_{\eta m} \right) + g^{bc} \left( \Gamma^i_{bd}\Gamma^d_{cm} + \Gamma^i_{b\eta }\Gamma^\eta_{cm} \right) \right] \mathcal{A}^m\\
    & = \left[ \Gamma^i_{\eta d}\Gamma^d_{\eta m} + g^{bc} \left( \Gamma^i_{bd}\Gamma^d_{cm} + \Gamma^i_{b\eta }\Gamma^\eta_{cm} \right) \right] \mathcal{A}^m\\
    & =\left[ \Gamma^i_{\eta d}\Gamma^d_{\eta e}\mathcal{A}^e + \Gamma^i_{\eta d}\Gamma^d_{\eta \eta }\mathcal{A}^\eta + g^{bc} \left( \Gamma^i_{bd}\Gamma^d_{ce}\mathcal{A}^e + \Gamma^i_{bd}\Gamma^d_{c\eta }\mathcal{A}^\eta + \Gamma^i_{b\eta }\Gamma^\eta_{ce}\mathcal{A}^e + \Gamma^i_{b\eta }\Gamma^\eta_{c\eta }\mathcal{A}^\eta \right) \right]\\
    &=  \left[ \Gamma^i_{\eta d}\Gamma^d_{\eta e}\mathcal{A}^e + g^{bc} \left( \Gamma^i_{bd}\Gamma^d_{ce}\mathcal{A}^e + \Gamma^i_{bd}\Gamma^d_{c\eta }\mathcal{A}^\eta + \Gamma^i_{b\eta }\Gamma^\eta_{ce}\mathcal{A}^e \right) \right].
\end{aligned}
\end{equation}
For the normal component, we get
\begin{equation}
    \begin{aligned}
    \sqrt{\Omega}\,T_3^\eta 
    & =  g^{bc} \left( \Gamma^\eta_{bd}\Gamma^d_{ce}\mathcal{A}^e + \Gamma^\eta_{bd}\Gamma^d_{c3}\mathcal{A}^\eta \right)\\
    &= g^{bc} \left( h_{bd} \Gamma^d_{ce}\mathcal{A}^e - h_{bd} h^d_c \mathcal{A}^\eta \right)\\
    &\approx g^{bc} h_{bd} \Gamma^d_{ce}\mathcal{A}^e\\
    &\approx h^c_d \Gamma^d_{ce}\mathcal{A}^e.
\end{aligned}
\end{equation}
The tangential part reads
\begin{equation}
    \begin{aligned}
   \sqrt{\Omega}\, T_3^a 
    &= \left[ \Gamma^a_{\eta d}\Gamma^d_{\eta e}\mathcal{A}^e + g^{bc} \left( \Gamma^a_{bd}\Gamma^d_{ce}\mathcal{A}^e + \Gamma^a_{bd}\Gamma^d_{c\eta }\mathcal{A}^\eta + \Gamma^a_{b\eta }\Gamma^\eta_{ce}\mathcal{A}^e \right) \right]\\
    & =\left[ (-h^a_d)(-h^d_e)\mathcal{A}^e + g^{bc} \left( \Gamma^a_{bd}\Gamma^d_{ce}\mathcal{A}^e + \Gamma^a_{bd}(-h^d_c)\mathcal{A}^\eta + (-h^a_b)(h_{ce})\mathcal{A}^e \right) \right]\\
    &\approx  g^{bc} \Gamma^a_{bd}\Gamma^d_{ce}\mathcal{A}^e - g^{bc} \Gamma^a_{bd} h^d_c \mathcal{A}^\eta\\
    &\approx g^{bc} \Gamma^a_{bd}\Gamma^d_{ce}\mathcal{A}^e - h^{bd} \Gamma^a_{bd} \mathcal{A}^\eta\\
    &\approx (\gamma^{bc} + 2\eta h^{bc}) \Gamma^a_{bd}\Gamma^d_{ce}\mathcal{A}^e - h^{bd} \Gamma^a_{bd} \mathcal{A}^\eta\\
    & \approx \gamma^{bc} \Gamma^a_{bd}\Gamma^d_{ce}\mathcal{A}^e + 2\eta h^{bc} \Gamma^a_{bd}\Gamma^d_{ce}\mathcal{A}^e - h^{bd} \Gamma^a_{bd} \mathcal{A}^\eta.
\end{aligned}
\end{equation}

\subsection{Final wave equation for $i=\eta$}
Putting all pieces together and dividing through by $\sqrt{\Omega}=\sqrt{1-2\eta H}$, which is nonzero except at a point $\eta=1/(2H)$ far away from the interface, yields the following wave equation for the normal vector field component
\begin{equation}
     \left[\Delta_\gamma + \partial_\eta^2 + k_0^2\epsilon(\eta)\right]\!\mathcal{A}^\eta + 2\eta\,h^{ab}\nabla_a\nabla_b\mathcal{A}^\eta+2 h^b_d \nabla_b \mathcal{A}^d=0,
\end{equation}
where we used 
\begin{equation}
   \sqrt{\Omega} \left(T_1^\eta + T_2^\eta + T_3^\eta\right)=  h^b_d \nabla_b \mathcal{A}^d + h^c_d \left( \partial_c \mathcal{A}^d + \Gamma^d_{ce}\mathcal{A}^e\right)=h^b_d \nabla_b \mathcal{A}^d +  h^c_d \nabla_c \mathcal{A}^d=2 h^b_d \nabla_b \mathcal{A}^d.
\end{equation}

\subsection{Final wave equation for $i=a$}
The three terms in term II of Eq.~\eqref{eq:app_wave_eq} for $i=a$, which we must add together, are
\begin{equation}
    \begin{aligned}
    &\sqrt{\Omega}\, T_1^a = - h^a_d \partial_\eta \mathcal{A}^d+ \frac{1}{\sqrt{\gamma}} \partial_b \left[ \sqrt{\gamma} \left( \gamma^{bc} \Gamma^a_{cd}\mathcal{A}^d - h^{ab} \mathcal{A}^\eta + 2\eta h^{bc} \Gamma^a_{cd}\mathcal{A}^d \right) \right],\\
    &\sqrt{\Omega}\,T_2^a = -h^a_d \partial_\eta \mathcal{A}^d +  \gamma^{bc} \Gamma^a_{bd} \partial_c \mathcal{A}^d + 2\eta h^{bc} \Gamma^a_{bd} \partial_c \mathcal{A}^d - h^{ac} \partial_c \mathcal{A}^\eta,\\
    &\sqrt{\Omega}\,T_\eta^a = \gamma^{bc} \Gamma^a_{bd}\Gamma^d_{ce}\mathcal{A}^e + 2\eta h^{bc} \Gamma^a_{bd}\Gamma^d_{ce}\mathcal{A}^e - h^{bd} \Gamma^a_{bd} \mathcal{A}^\eta.
\end{aligned}
\end{equation}
We define three groups: $S_1$ contains all terms with the normal derivative $\partial_\eta \mathcal{A}^d$, $S_2$ collects all terms containing the normal field component $\mathcal{A}^\eta$ or its derivatives, and $S_3$ contains the tangential field $\mathcal{A}^d$ and its surface derivatives. The first group is written as
\begin{equation}
    S_1 = - 2 h^a_d \partial_\eta \mathcal{A}^d.
\end{equation}
The second group is
\begin{equation}
    \begin{aligned}
    S_2& = \frac{1}{\sqrt{\gamma}} \partial_b \left[ \sqrt{\gamma} (- h^{ab} \mathcal{A}^\eta) \right] - h^{ac} \partial_c \mathcal{A}^\eta  - h^{bd} \Gamma^a_{bd} \mathcal{A}^\eta\\
    &= \left[ \frac{1}{\sqrt{\gamma}} (\partial_b \sqrt{\gamma}) (- h^{ab} \mathcal{A}^\eta) - \partial_b (h^{ab} \mathcal{A}^\eta) \right] - h^{ac} \partial_c \mathcal{A}^\eta  - h^{bd} \Gamma^a_{bd} \mathcal{A}^\eta\\
    &= \left[ - \Gamma^c_{bc} h^{ab} \mathcal{A}^\eta - (\partial_b h^{ab}) \mathcal{A}^\eta - h^{ab} (\partial_b \mathcal{A}^\eta) \right] - h^{ac} \partial_c \mathcal{A}^\eta  - h^{bd} \Gamma^a_{bd} \mathcal{A}^\eta\\
    &=  - \Gamma^c_{bc}h^{ab}  \mathcal{A}^\eta- (\partial_b h^{ab}) \mathcal{A}^\eta - 2h^{ab} (\partial_b \mathcal{A}^\eta) - h^{bc} \Gamma^a_{bc} \mathcal{A}^\eta\\
    &=- 2 h^{ab} \partial_b \mathcal{A}^\eta - \left[ \partial_b h^{ab} + \Gamma^c_{bc} h^{ab} + \Gamma^a_{bc} h^{bc} \right] \mathcal{A}^\eta\\
    &=- 2 h^{ab} \partial_b \mathcal{A}^\eta - \left(\bar \nabla_b h^{ab}\right)\mathcal{A}^\eta\\
    &\approx - 2 h^{ab} \partial_b \mathcal{A}^\eta,
\end{aligned}
\end{equation}
since $\nabla_b h^{ab}\approx 0$ (from the Mainardi-Codazzi equations). The third group is
\begin{equation}
    \begin{aligned}
    S_3&= \frac{1}{\sqrt{\gamma}} \partial_b \left[ \sqrt{\gamma} \left( \gamma^{bc} \Gamma^a_{cd}\mathcal{A}^d + 2\eta h^{bc} \Gamma^a_{cd}\mathcal{A}^d \right) \right]\\
    &\quad + \gamma^{bc} \Gamma^a_{bd} \partial_c \mathcal{A}^d + 2\eta h^{bc} \Gamma^a_{bd} \partial_c \mathcal{A}^d\\
    &\quad + \gamma^{bc} \Gamma^a_{bd}\Gamma^d_{ce}\mathcal{A}^e + 2\eta h^{bc} \Gamma^a_{bd}\Gamma^d_{ce}\mathcal{A}^e\\
    &=  \frac{1}{\sqrt{\gamma}} \partial_b \left[ \sqrt{\gamma} \left( \gamma^{bc} + 2\eta h^{bc} \right) \Gamma^a_{cd}\mathcal{A}^d \right]
    + \left( \gamma^{bc} + 2\eta h^{bc} \right) \Gamma^a_{bd} \partial_c \mathcal{A}^d+\left( \gamma^{bc} + 2\eta h^{bc} \right) \Gamma^a_{bd} \Gamma^d_{ce} \mathcal{A}^e\\
    &=  \frac{1}{\sqrt{\gamma}} \partial_b \left( \sqrt{\gamma} g^{bc} \Gamma^a_{cd}\mathcal{A}^d \right)
    + g^{bc} \Gamma^a_{bd} \left(\partial_c \mathcal{A}^d+ \Gamma^d_{ce} \mathcal{A}^e\right)\\
    &=  \frac{1}{\sqrt{\gamma}} \partial_b \left( \sqrt{\gamma} g^{bc} \Gamma^a_{cd}\mathcal{A}^d \right)
    + g^{bc} \Gamma^a_{cd} \nabla_b \mathcal{A}^d.
\end{aligned}
\end{equation}
Therefore, we obtain
\begin{equation}
    \sqrt{\Omega}\left(T_1^a + T_2^a + T_3^a\right)= - 2 h^a_d \partial_\eta \mathcal{A}^d - 2 h^{ab} \partial_b \mathcal{A}^\eta + C^a,
\end{equation}
with the short-hand notation
\begin{equation}
    C^a \equiv \frac{1}{\sqrt{\gamma}} \partial_b \left( \sqrt{\gamma} g^{bc} \Gamma^a_{cd}\mathcal{A}^d \right) + g^{bc} \Gamma^a_{cd} \nabla_b \mathcal{A}^d,
\end{equation}
since $\bar{R}^a_{\ cdb}=\mathcal{O}(K)$.
After again dividing through by $\sqrt{\Omega}$, the wave equation for the tangential vector field component can then be written as
\begin{equation}\label{eq:wave_equation_tangential}
   \left[\Delta_\gamma + \partial_\eta^2 + k_0^2\epsilon(\eta)\right]\!\mathcal{A}^a + 2\eta\,h^{bc}\nabla_b\nabla_c\mathcal{A}^a- 2 h^a_d \partial_\eta \mathcal{A}^d  + C^a=2 h^{ab} \partial_b \mathcal{A}^\eta,
\end{equation}
which is the same expanded form as written in Ref.~\cite{lai2018electromagnetic}.

Eq.~\eqref{eq:wave_equation_tangential} shows that the tangential components experience the normal vector field $\mathcal{A}^\eta$ as a source term in the presence of curvature. Since we use an expansion in orders of curvature in Section~\ref{supp:derivation_final_equation}, we must take care of the scaling of $C^a$. As explained in these sections, we can drop it if it scales at least as $h^{ab}\sim H$. With $g^{ab}=\gamma^{ab}+2\eta h^{ab}$, we split $C^a$ into four parts as
\begin{equation}
    C^a = \frac{1}{\sqrt{\gamma}} \partial_b \left( \sqrt{\gamma} \gamma^{bc} \Gamma^a_{cd}\mathcal{A}^d \right) + \gamma^{bc} \Gamma^a_{cd} \nabla_b \mathcal{A}^d + \frac{2\eta}{\sqrt{\gamma}} \partial_b \left( \sqrt{\gamma} h^{bc} \Gamma^a_{cd}\mathcal{A}^d \right) + 2\eta h^{bc} \Gamma^a_{cd} \nabla_b \mathcal{A}^d.
\end{equation}
The last two terms will be dropped as these are at least $\mathcal{O}(H)$. Using a Monge patch parametrization $\mathbf{r}(x^1,x^2)=\left(x^1,x^2, f(x^1,x^2) \right)$ with a height function $f(x^1, x^2)$, one can show that the intrinsic surface Christoffel symbols scale as~\cite{deserno2004notes}
\begin{equation}
    \Gamma^{a}_{bc}=\mathcal{O}{\alpha^2} \qquad \implies \qquad \partial_c \gamma^{ab}=-\left(\gamma^{bd}\Gamma^{a}_{dc}+ \gamma^{ad}\Gamma^{b}_{dc}\right)=\mathcal{O}{\alpha^2}.
\end{equation}
Therefore, $C^a$ scales as $\mathcal{O}{\alpha^2}$.

\section{Maxwell boundary matching conditions}\label{supp:curved_boundary_conditions}
In the Lorenz gauge, the scalar potential is given by $\phi = (i \omega \epsilon \mu)^{-1}\nabla_j A^j$. Assuming no surface charges or dipole layers, both the tangential electric field and the scalar potential $\phi$ must be continuous across the interface. For a non-magnetic system ($\mu_d = \mu_m$), this requires the continuity of $\epsilon^{-1} \nabla_j A^j$, which means
\begin{equation}
     \frac{1}{\epsilon} \nabla_j A^j \Bigg|_{\eta=0^+} =  \frac{1}{\epsilon} \nabla_j A^j \Bigg|_{\eta=0^-}.
\end{equation}
Expanding the covariant divergence $\nabla_j A^j = \frac{1}{\sqrt{g}} \partial_i (\sqrt{g} A^i)$ into normal ($\eta$) and tangential ($a=1,2$) components at the interface ($\eta=0$) yields
\begin{equation}\label{eq:divergence_unscaled}
    \nabla_j A^j = \partial_\eta A^\eta - 2H A^\eta + \nabla_a A^a,
\end{equation}
for which we used the expansion of $\sqrt{g} = \Omega \sqrt{\gamma}$ with the factor $\Omega \approx 1 - 2H\eta$ near the interface, resulting in $\partial_\eta \Omega|_{\eta=0} = -2H$. To express this in terms of the rescaled field $\mathcal{A}^i = \sqrt{\Omega}\,A^i$ required for the correct volume measure~\cite{da1981quantum, batz2008linear, carmi2025photon}, we evaluate the normal derivative of the physical field at $\eta=0$. Since $\sqrt{\Omega}\big|_{\eta=0} = 1$ and $\partial_\eta \sqrt{\Omega}\big|_{\eta=0} = -H$, we get
\begin{equation}
    \partial_\eta A^\eta\bigg|_{\eta=0} = \partial_\eta \left( \frac{\mathcal{A}^\eta}{\sqrt{\Omega}} \right)\Bigg|_{\eta=0} = \partial_\eta \mathcal{A}^\eta + H \mathcal{A}^\eta.
\end{equation}
Substituting this into Eq.~\eqref{eq:divergence_unscaled} gives
\begin{equation}
    \nabla_j A^j\bigg|_{\eta=0} = \partial_\eta \mathcal{A}^\eta - H \mathcal{A}^\eta + \nabla_a \mathcal{A}^a.
\end{equation}

As the physical field $\mathbf{A}$ is continuous across the interface, the rescaled fields $\mathcal{A}^\eta$ and $\mathcal{A}^a$, as well as the tangential divergence $\nabla_a \mathcal{A}^a$ are also continuous. The  continuity condition then reads
\begin{equation}
    \frac{1}{\epsilon_d} \left( \partial_\eta \mathcal{A}_d^\eta - H \mathcal{A}_d^\eta + \nabla_a \mathcal{A}_d^a \right) 
    = 
    \frac{1}{\epsilon_m} \left( \partial_\eta \mathcal{A}_m^\eta - H \mathcal{A}_m^\eta + \nabla_a \mathcal{A}_m^a \right),
\end{equation}
where the lower index means dielectric $d$ for $\eta=0^+$ and metal $m$ for $\eta=0^-$. Rearranging the terms yields the curvature-modified boundary matching condition used in the main text as Eq.~(2).

\section{Derivation of the effective surface wave equation}\label{supp:derivation_final_equation}

Starting from the separable flat-interface SPP mode, see Eq.~\eqref{eq:separation_Az}, we write
\begin{equation}
    \mathcal{A}_i^\eta=f_i(\eta)\psi(q_{\parallel}),
\end{equation}
and expand both parts in orders of $\alpha$ as
\begin{equation}
    \begin{aligned}
        f_i(\eta)=f_i^{(0)}(\eta)+\alpha f_i^{(1)}(\eta)+\mathcal{O}(\alpha^2),\\
        \psi(q_{\parallel})=\psi^{(0)}(q_{\parallel})+\alpha \psi^{(1)}(q_{\parallel})+\mathcal{O}(\alpha^2),
    \end{aligned}
\end{equation}
with $f_i^{(0)}(\eta)=e^{-\kappa_i|\eta|}$.
A first ansatz for $f_i^{(1)}(\eta)$ would be an exponential function with correction terms only for the decay parameter $\kappa_i$ as $ e^{-(\kappa_i -\alpha B_i)|\eta|}\approx \alpha e^{-\kappa_i|\eta|}\left(B_i \eta \right) +\mathcal{O}{\alpha^2}$. However, by using this expression, we could not close the system of equations, and we therefore chose the more general ansatz
\begin{equation}\label{eq:ansatz_A_eta}
    f_i^{(1)}(\eta)=e^{-\kappa_i|\eta|}\left( A_i \eta^2 + B_i \eta + C_i\right).
\end{equation}
This yields
\begin{equation} \label{eq:app_ansatz_eff_surface_wavefct}
    \begin{aligned}
        \mathcal{A}_i^\eta &= \left( f_i^{(0)}(\eta) + \alpha f_i^{(1)}(\eta) \right) \left( \psi^{(0)} + \alpha \psi^{(1)} \right)\\
        & = f_i^{(0)}(\eta) \psi^{(0)} + \alpha \left[ f_i^{(0)}(\eta) \psi^{(1)} + f_i^{(1)}(\eta) \psi^{(0)} \right] +\mathcal{O}(\alpha^2).
    \end{aligned}
\end{equation}

\subsection{Dielectric domain}
We define 
\begin{equation}\label{eq:L_flat_alpha_squared}
\hat{L} \equiv \underbrace{ \Delta_{\rm flat} + \partial_\eta^2 + k_0^2 \epsilon(\eta) }_{\hat{L}^{(0)}} + \alpha \underbrace{ 2\eta \mathcal{D}^{(1)}_h }_{\hat{L}^{(1)}},
\end{equation}
with
\begin{equation}
\mathcal{D}^{(1)}_h \equiv H^{(1)}\Delta_{\rm flat} + \sigma^{ab\,(1)}\partial_a\partial_b.
\end{equation}
For the dielectric region, the ansatz in Eq.~\eqref{eq:app_ansatz_eff_surface_wavefct} reads
\begin{equation}
 \mathcal{A}_d^\eta = \underbrace{ e^{-\kappa_d \eta} \psi^{(0)} }_{\mathcal{A}_d^{\eta\,(0)}} + \alpha \underbrace{ e^{-\kappa_d \eta} \left[ \psi^{(1)} + (A_d \eta^2 + B_d \eta + C_d) \psi^{(0)} \right] }_{\mathcal{A}_d^{\eta\,(1)}}.
\end{equation}
The equation $\hat{L} \mathcal{A}_d^\eta = 0$, with $\hat{L}$ as in Eq.~\eqref{eq:L_flat_alpha_squared}, must now be satisfied at each order in $\alpha$.\\[0.5cm]
\textbf{Order $\alpha^0$}:\\
At zero order, we get
\begin{equation}
    \hat{L}^{(0)}\mathcal{A}_d^{\eta\,(0)}=\left[\Delta_{\rm flat} + \partial_\eta^2 + k_0^2 \epsilon(\eta)  \right]e^{-\kappa_d \eta} \psi^{(0)}=0,
\end{equation}
which can be written as
\begin{equation}
    \psi^{(0)}\left[\partial_\eta^2 + k_0^2 \epsilon(\eta)  \right]e^{-\kappa_d \eta}+e^{-\kappa_d \eta}\left(\Delta_{\rm flat} \psi^{(0)}\right)=0.
\end{equation}
Applying the second-order derivative on the exponential and using that $\kappa^2_d=k^2_{\rm spp}-k^2_0\epsilon_d$, see Eq.~\eqref{eq:kappa_wavenumber}, yields, after factoring out the remaining exponential,
\begin{equation}
    \Delta_{\rm flat} \psi^{(0)}=-k^2_{\rm spp}\psi^{(0)}.
\end{equation}
This is simply the SPP Helmholtz equation for a flat interface, cf.~Section~\ref{sec:flat_spp_and_gauge}.\\[0.5cm]
\textbf{Order $\alpha^1$}:\\
\begin{equation}
 [\Delta_{\gamma}+2\eta\,\mathcal{D}_h+\partial_{\eta}^{2}+k_{0}^{2}\epsilon]\mathcal{A}^{\eta}+2 h^a_b \nabla_a \mathcal{A}^b=0,
\end{equation}
At first order, we obtain the inhomogeneous differential equation
\begin{equation}
    \hat{L}^{(0)} \mathcal{A}_d^{\eta\,(1)} + \hat{L}^{(1)} \mathcal{A}_d^{\eta\,(0)} +\mathcal{M}_a^{(1)} \mathcal{A}_d^{a\,(0)} = 0,
\end{equation}
with 
\begin{equation}
    \mathcal{M}_a^{(1)}\equiv 2 h^b_a \nabla_b. 
\end{equation}
First, we note again that the tangential fields for the flat interface vanish, i.e., $\mathcal{A}_d^{d\,(0)}=0$, leaving
\begin{equation}
     \hat{L}^{(0)} \mathcal{A}_d^{\eta\,(1)} + \hat{L}^{(1)} \mathcal{A}_d^{\eta\,(0)}=0,
\end{equation}
which is written out as
\begin{equation}
    \left[ \Delta_{\rm flat} + \partial_\eta^2 + k_0^2 \epsilon_d \right] \left( e^{-\kappa_d \eta} \Big[ \psi^{(1)} + (A_d \eta^2 + B_d \eta + C_d) \psi^{(0)} \Big] \right) + 2\eta \mathcal{D}^{(1)}_h \left( e^{-\kappa_d \eta} \psi^{(0)} \right) = 0.
\end{equation}

First, we evaluate the geometric source term. Because the operator $\mathcal{D}^{(1)}_h = H^{(1)} \Delta_{\rm flat} + \sigma^{ab\,(1)} \partial_a \partial_b$ consists only of surface derivatives ($\partial_a$), it commutes with the normal coordinate $\eta$ and the exponential $e^{-\kappa_d \eta}$, and acts only on the surface envelope $\psi^{(0)}$ as
\begin{equation}
    2\eta \mathcal{D}^{(1)}_h \left( e^{-\kappa_d \eta} \psi^{(0)} \right) = 2\eta e^{-\kappa_d \eta} \mathcal{D}^{(1)}_h \psi^{(0)}.
\end{equation}
Second, we evaluate the action of the flat surface Laplacian $\Delta_{\rm flat}$ on our $\mathcal{O}(\alpha)$ field ansatz. Like $\mathcal{D}^{(1)}_h$, this operator only acts on the surface functions $\psi^{(0)}$ and $\psi^{(1)}$ as
\begin{equation}
    \begin{aligned}
    \Delta_{\rm flat} \left( e^{-\kappa_d \eta} \Big[ \psi^{(1)} + (A_d \eta^2 + B_d \eta + C_d) \psi^{(0)} \Big] \right)
    &=
    e^{-\kappa_d \eta} \Delta_{\rm flat} \psi^{(1)} + e^{-\kappa_d \eta} (A_d \eta^2 + B_d \eta + C_d) \Delta_{\rm flat} \psi^{(0)}\\
    &=
    e^{-\kappa_d \eta} \left[ \Delta_{\rm flat} \psi^{(1)} - k_{\text{spp}}^2 (A_d \eta^2 + B_d \eta + C_d) \psi^{(0)} \right]
\end{aligned}
\end{equation}
for which we have used the zero-order result $\Delta_{\rm flat} \psi^{(0)} = -k_{\text{spp}}^2 \psi^{(0)}$ and the approximation $\Delta_{\rm flat}A_d\approx 0$, which will retrospectively correspond to $\partial_a H \approx 0$ (our slowly changing curvature assumption). Let us temporarily write the polynomial as $P(\eta) = A_d \eta^2 + B_d \eta + C_d$. The second-order field then reads
\begin{equation}
    \mathcal{A}_d^{\eta\,(1)} = e^{-\kappa_d \eta} \Big[ \psi^{(1)} + P(\eta) \psi^{(0)} \Big].
\end{equation}
We take the first normal derivative $\partial_\eta$ using the product rule and get
\begin{equation}
    \partial_\eta \mathcal{A}_d^{\eta\,(1)} = -\kappa_d e^{-\kappa_d \eta} \Big[ \psi^{(1)} + P(\eta) \psi^{(0)} \Big] + e^{-\kappa_d \eta} \Big[ P'(\eta) \psi^{(0)} \Big].
\end{equation}
Now we apply the product rule again to find the second derivative as
\begin{equation}
    \begin{aligned}
    \partial_\eta^2 \mathcal{A}_d^{\eta\,(1)} 
    &=
    \kappa_d^2 e^{-\kappa_d \eta} \Big[ \psi^{(1)} + P(\eta) \psi^{(0)} \Big] - \kappa_d e^{-\kappa_d \eta} \Big[ P'(\eta) \psi^{(0)} \Big] - \kappa_d e^{-\kappa_d \eta} \Big[ P'(\eta) \psi^{(0)} \Big] + e^{-\kappa_d \eta} \Big[ P''(\eta) \psi^{(0)} \Big]\\
    &=
    e^{-\kappa_d \eta} \left[ \kappa_d^2 \Big( \psi^{(1)} + P(\eta) \psi^{(0)} \Big) - 2\kappa_d P'(\eta) \psi^{(0)} + P''(\eta) \psi^{(0)} \right].
\end{aligned}
\end{equation}
The derivatives of the polynomial are
\begin{equation}
    P'(\eta) = 2A_d \eta + B_d, \qquad P''(\eta) = 2A_d.
\end{equation}
Substituting these back into our second derivative and adding the $k_0^2 \epsilon_d$ term from the operator, we find
\begin{equation}
    (\partial_\eta^2 + k_0^2 \epsilon_d) \mathcal{A}_d^{\eta\,(1)} = e^{-\kappa_d \eta} \left[ (\kappa_d^2 + k_0^2 \epsilon_d) \Big( \psi^{(1)} + (A_d \eta^2 + B_d \eta + C_d) \psi^{(0)} \Big) - 2\kappa_d (2A_d \eta + B_d) \psi^{(0)} + 2A_d \psi^{(0)} \right].
\end{equation}
The full action of the zero-order operator $\hat{L}^{(0)}$ on the second-order field $\mathcal{A}_d^{\eta\,(1)}$ is then
\begin{equation}
    \begin{aligned}
    \hat{L}^{(0)} \mathcal{A}_d^{\eta\,(1)} =
    e^{-\kappa_d \eta} \bigg[ &\Delta_{\rm flat} \psi^{(1)} - k_{\text{spp}}^2 \Big( A_d \eta^2 + B_d \eta + C_d \Big) \psi^{(0)}\\
    &+
    (\kappa_d^2 + k_0^2 \epsilon_d) \Big( \psi^{(1)} + (A_d \eta^2 + B_d \eta + C_d) \psi^{(0)} \Big) - 2\kappa_d (2A_d \eta + B_d) \psi^{(0)} + 2A_d \psi^{(0)} \bigg].
\end{aligned}
\end{equation}
The two appearing full polynomials cancel, since from Eq.~\eqref{eq:kappa_wavenumber} we have $\kappa_d^2 + k_0^2 \epsilon_d = k_{\text{spp}}^2$, leaving 
\begin{equation}
    \hat{L}^{(0)} \mathcal{A}_d^{\eta\,(1)} = e^{-\kappa_d \eta} \bigg[ (\Delta_{\rm flat} + k_{\text{spp}}^2) \psi^{(1)} + 2A_d \psi^{(0)} - 2\kappa_d B_d \psi^{(0)} - 4\kappa_d A_d \eta \psi^{(0)} \bigg].
\end{equation}
We can now also add $2\eta e^{-\kappa_d \eta} \mathcal{D}^{(1)}_h \psi^{(0)}$ and divide through by the exponential $e^{-\kappa_d \eta}$, which is non-zero in the dielectric domain, to get
\begin{equation}
    (\Delta_{\rm flat} + k_{\text{spp}}^2) \psi^{(1)} + (2A_d - 2\kappa_d B_d) \psi^{(0)} + \eta \Big[2\mathcal{D}^{(1)}_h  -4\kappa_d A_d \Big]\psi^{(0)} = 0.
\end{equation}
Since this differential equation must hold true for every arbitrary value of $\eta > 0$ within the dielectric domain, the coefficients of each linearly independent power of $\eta$ must vanish. This yields
\begin{equation}
    2\mathcal{D}^{(1)}_h\psi^{(0)}  -4\kappa_d A_d\psi^{(0)} = 0,
\end{equation}
which is solved as
\begin{equation}
    A_d \psi^{(0)} = \frac{1}{2\kappa_d} \mathcal{D}^{(1)}_h \psi^{(0)},
\end{equation}
and 
\begin{equation}
    (\Delta_{\rm flat} + k_{\text{spp}}^2) \psi^{(1)} + 2A_d \psi^{(0)} - 2\kappa_d B_d \psi^{(0)} = 0,
\end{equation}
which is solved as
\begin{equation}
    B_d \psi^{(0)} = \frac{1}{2\kappa_d} (\Delta_{\rm flat} + k_{\text{spp}}^2) \psi^{(1)} + \frac{1}{2\kappa_d^2} \mathcal{D}^{(1)}_h \psi^{(0)}.
\end{equation}

\subsection{Metal domain}
In the metal, the field decays as $e^{\kappa_m \eta}$, such that the ansatz in Eq.~\eqref{eq:app_ansatz_eff_surface_wavefct} reads
\begin{equation}
 \mathcal{A}_m^\eta = \underbrace{ e^{\kappa_m \eta} \psi^{(0)} }_{\mathcal{A}_m^{\eta\,(0)}} + \alpha \underbrace{ e^{\kappa_m \eta} \left[ \psi^{(1)} + (A_m \eta^2 + B_m \eta + C_m) \psi^{(0)} \right] }_{\mathcal{A}_m^{\eta\,(1)}}.
\end{equation}
We perform exactly the same steps as in the dielectric. This yields similarly
\begin{equation}
    A_m \psi^{(0)} = -\frac{1}{2\kappa_m} \mathcal{D}^{(1)}_h \psi^{(0)},
\end{equation}
and
\begin{equation}
    B_m \psi^{(0)} = -\frac{1}{2\kappa_m} (\Delta_{\rm flat} + k_{\text{spp}}^2) \psi^{(1)} + \frac{1}{2\kappa_m^2} \mathcal{D}^{(1)}_h \psi^{(0)}.
\end{equation}

\subsection{Evaluating the fields at the boundary}
The fields at $\eta=0$ are evaluated as
\begin{equation}
    \mathcal{A}_d^{\eta\,(1)} \Big|_{\eta=0} = \psi^{(1)} + C_d \psi^{(0)}, \qquad
    \mathcal{A}_m^{\eta\,(1)} \Big|_{\eta=0} = \psi^{(1)} + C_m \psi^{(0)}.
\end{equation}
The normal field component must be continuous across the interface due to our gauge choice, and we obtain
\begin{equation}
    \psi^{(1)} + C_d \psi^{(0)} = \psi^{(1)} + C_m \psi^{(0)},
\end{equation}
which means $C_d = C_m\equiv C$. Let us recall
\begin{equation}
    \partial_\eta \mathcal{A}_d^{\eta\,(1)} = -\kappa_d e^{-\kappa_d \eta} \Big[ \psi^{(1)} + (A_d \eta^2 + B_d \eta + C_d) \psi^{(0)} \Big] + e^{-\kappa_d \eta} \Big[ (2A_d \eta + B_d) \psi^{(0)} \Big],
\end{equation}
and 
\begin{equation}
    \partial_\eta \mathcal{A}_m^{\eta\,(1)} = \kappa_m e^{\kappa_m \eta} \Big[ \psi^{(1)} + (A_m \eta^2 + B_m \eta + C_m) \psi^{(0)} \Big] + e^{\kappa_m \eta} \Big[ (2A_m \eta + B_m) \psi^{(0)} \Big],
\end{equation}
which, at the boundary, take the values
\begin{equation}
    \partial_\eta \mathcal{A}_d^{\eta\,(1)} \Big|_{\eta=0} = -\kappa_d \psi^{(1)} - \kappa_d C \psi^{(0)} + \frac{1}{2\kappa_d} (\Delta_{\rm flat} + k_{\text{spp}}^2) \psi^{(1)} + \frac{1}{2\kappa_d^2} \mathcal{D}^{(1)}_h \psi^{(0)} ,
\end{equation}
and
\begin{equation}
    \partial_\eta \mathcal{A}_m^{\eta\,(1)} \Big|_{\eta=0} = \kappa_m \psi^{(1)} + \kappa_m C \psi^{(0)} -\frac{1}{2\kappa_m} (\Delta_{\rm flat} + k_{\text{spp}}^2) \psi^{(1)} + \frac{1}{2\kappa_m^2} \mathcal{D}^{(1)}_h \psi^{(0)},
\end{equation}
where we already inserted the above derived expressions for $A_{m,d}$ and $B_{m,d}$. The tangential divergence, see Section~\ref{supp:derivation_tangential_divergence}, evaluated at the boundary takes the value
\begin{equation}
    \left(\nabla_a \mathcal{A}_d^{a}\right)^{(1)}\Big|_{\eta=0} 
    =
    \left(\nabla_a \mathcal{A}_m^{a}\right)^{(1)}\Big|_{\eta=0} 
    =
    -\frac{1}{k^2_{\rm spp}}\mathcal{D}^{(1)}_h \psi^{(0)}
    =
     -\frac{1}{k^2_{\rm spp}}\left(-H^{(1)}k^2_{\rm spp} + \sigma^{ab\,(1)}\partial_a\partial_b\right) \psi^{(0)}.
\end{equation}

\subsection{Boundary matching condition}
The boundary matching condition taking into account the curvature is derived in Section~\ref{supp:curved_boundary_conditions},  and reads at first order in $\alpha$
\begin{equation}
    \frac{1}{\epsilon_d} \bigg[ \left(\nabla_a \mathcal{A}_d^{a}\right)^{(1)} + \partial_\eta \mathcal{A}_d^{\eta\,(1)} - H^{(1)} \mathcal{A}_d^{\eta\,(0)} \bigg] 
    = 
    \frac{1}{\epsilon_m} \bigg[ \left(\nabla_a \mathcal{A}_m^{a}\right)^{(1)} + \partial_\eta \mathcal{A}_m^{\eta\,(1)} - H^{(1)} \mathcal{A}_m^{\eta\,(0)} \bigg].
\end{equation}
The left-hand side (LHS) of this equation writes
\begin{equation}
\text{LHS} = \frac{1}{\epsilon_d} \bigg[ -\frac{1}{k_{\text{spp}}^2} \mathcal{D}^{(1)}_h \psi^{(0)} \underbrace{ -\kappa_d \psi^{(1)} - \kappa_d C \psi^{(0)} + \frac{1}{2\kappa_d} (\Delta_{\rm flat} + k_{\text{spp}}^2) \psi^{(1)} + \frac{1}{2\kappa_d^2} \mathcal{D}^{(1)}_h \psi^{(0)} }_{\partial_\eta \mathcal{A}_d^{\eta\,(1)}} - H^{(1)} \psi^{(0)} \bigg],
\end{equation}
while the RHS writes
\begin{equation}
    \text{RHS} = \frac{1}{\epsilon_m} \bigg[ -\frac{1}{k_{\text{spp}}^2} \mathcal{D}^{(1)}_h \psi^{(0)} \underbrace{ +\kappa_m \psi^{(1)} + \kappa_m C \psi^{(0)} - \frac{1}{2\kappa_m} (\Delta_{\rm flat} + k_{\text{spp}}^2) \psi^{(1)} + \frac{1}{2\kappa_m^2} \mathcal{D}^{(1)}_h \psi^{(0)} }_{\partial_\eta \mathcal{A}_m^{\eta\,(1)}} - H^{(1)} \psi^{(0)} \bigg].
\end{equation}
We now move the terms with $C$ and the bare $\psi^{(1)}$ terms to the left side. These are
\begin{equation} \label{eq:app_BC_on_psi0}
    -\frac{\kappa_d}{\epsilon_d} C \psi^{(0)} - \frac{\kappa_m}{\epsilon_m} C \psi^{(0)} = - \left( \frac{\kappa_d}{\epsilon_d} + \frac{\kappa_m}{\epsilon_m} \right) C \psi^{(0)},
\end{equation}
and
\begin{equation} \label{eq:app_BC_on_psi1}
    -\frac{\kappa_d}{\epsilon_d} \psi^{(1)} - \frac{\kappa_m}{\epsilon_m} \psi^{(1)} = - \left( \frac{\kappa_d}{\epsilon_d} + \frac{\kappa_m}{\epsilon_m} \right) \psi^{(1)}.
\end{equation}
Since $\kappa_d/\epsilon_d+\kappa_m/\epsilon_m=0$, both Eqs.~\eqref{eq:app_BC_on_psi0} and \eqref{eq:app_BC_on_psi1} vanish. Moving the remaining $\psi^{(1)}$ terms to the left and grouping the $\psi^{0}$ terms on the right, we arrive at
\begin{equation}
    \left( \frac{1}{2\epsilon_d \kappa_d} + \frac{1}{2\epsilon_m \kappa_m} \right) (\Delta_{\text{flat}} + k_{\text{spp}}^2) \psi^{(1)} 
    =
    \left[ \frac{\mathcal{D}^{(1)}_h}{\epsilon_m} \left( \frac{1}{2\kappa_m^2} - \frac{1}{k_{\text{spp}}^2} \right) - \frac{\mathcal{D}^{(1)}_h}{\epsilon_d} \left( \frac{1}{2\kappa_d^2} - \frac{1}{k_{\text{spp}}^2} \right) + H^{(1)} \left( \frac{1}{\epsilon_d} - \frac{1}{\epsilon_m} \right) \right]  \psi^{(0)}.
\end{equation}
To isolate the 2D wave operator $(\Delta_{\text{flat}} + k_{\text{spp}}^2) \psi^{(1)}$, we divide the entire RHS by the corresponding prefactor and define the resulting RHS as the geometric potential operator $V_{\rm geom}$, such that
\begin{equation}
    (\Delta_{\rm flat} + k_{\text{spp}}^2) \psi^{(1)} = V_{\rm geom} \psi^{(0)}.
\end{equation}

We now have two differential equations: 
\begin{itemize}
    \item $\mathcal{O}(\alpha^0)$: \quad $(\Delta_{\rm flat} + k_{\text{spp}}^2) \psi^{(0)} = 0$, and
    \item $\mathcal{O}(\alpha^1)$: \quad $(\Delta_{\rm flat} + k_{\text{spp}}^2) \psi^{(1)} = V_{\rm geom} \psi^{(0)}$.
\end{itemize}
Let us use these results to reconstruct the equation for the full envelope $\psi = \psi^{(0)} + \alpha \psi^{(1)}$. We therefore add the two equations and use $\psi^{(0)}=\psi-\alpha\psi^{(1)}$ to get
\begin{equation}
    (\Delta_{\rm flat} + k_{\text{spp}}^2) \big( \psi^{(0)} + \alpha \psi^{(1)} \big) = \alpha V_{\rm geom} \psi-\alpha^2V_{\rm geom} \psi^{(1)},
\end{equation}
and discard the $\alpha^2$ term, such that
\begin{equation}\label{eq:preliminary_result_delta_flat}
    \left[ \Delta_{\rm flat} + k_{\text{spp}}^2 - \alpha V_{\rm geom} \right] \psi = 0.
\end{equation}
We now have to calculate the specific expression for the geometric potential, which is
\begin{equation}
    V_{\rm geom}=\left( \frac{1}{2\epsilon_d \kappa_d} + \frac{1}{2\epsilon_m \kappa_m} \right)^{-1}\left[ \frac{\mathcal{D}^{(1)}_h}{\epsilon_m} \left( \frac{1}{2\kappa_m^2} - \frac{1}{k_{\text{spp}}^2} \right) - \frac{\mathcal{D}^{(1)}_h}{\epsilon_d} \left( \frac{1}{2\kappa_d^2} - \frac{1}{k_{\text{spp}}^2} \right) + H^{(1)} \left( \frac{1}{\epsilon_d} - \frac{1}{\epsilon_m} \right) \right].
\end{equation}
Simplifying the prefactors yields after an algebraic calculation
\begin{equation}
    V_{\rm geom}= - \frac{k_0 (\epsilon_d^2 + \epsilon_d \epsilon_m + \epsilon_m^2)}{(\epsilon_d + \epsilon_m) \sqrt{-(\epsilon_d + \epsilon_m)}}H^{(1)}+ \frac{\epsilon_d^2 + 3\epsilon_d \epsilon_m + \epsilon_m^2}{k_0 \epsilon_d \epsilon_m \sqrt{-(\epsilon_d + \epsilon_m)}}\sigma^{ab\,(1)}\partial_a\partial_b.
\end{equation}
Finally, we can write the effective wave equation for the curvature-modified TM SPP envelope $\psi$ on the metal-dielectric interface as
\begin{equation}\label{eq:final_wave_equation_appendix}
    \left[ \Delta_{\gamma} + k_{\text{spp}}^2 
    +
    \frac{k_0 (\epsilon_d^2 + \epsilon_d \epsilon_m + \epsilon_m^2)}{(\epsilon_d + \epsilon_m) \sqrt{-(\epsilon_d + \epsilon_m)}}H
    -
\frac{\epsilon_d^2 + 3\epsilon_d \epsilon_m + \epsilon_m^2}{k_0 \epsilon_d \epsilon_m \sqrt{-(\epsilon_d + \epsilon_m)}}\sigma^{ab}\nabla_a\nabla_b
    \right] \psi = 0,
\end{equation}
which corresponds to Eq.~(3) in the main text. 

While for the derivation of $V_H$ and $V_\sigma$ we only needed the flat-interface derivative operators, we finally restored their covariant versions $\Delta_{\mathrm{flat}} \to \Delta_\gamma$ and $\partial \to \nabla$, which yields an equation applicable also to closed surfaces without the need to glue together several local patches, and which is invariant under coordinate transformations. Since intrinsic curvature corrections scale as the Gaussian curvature $K \sim \mathcal{O}(\alpha^2)$, this does not affect any $\mathcal{O}{\alpha}$ result emerging from Eq.~(3). As a demonstration, we show in Fig.~\ref{fig:collective_supp}, Section~\ref{supp:greens_interaction_full}, the collective eigenvalues of a ring of emitters placed above a metal sphere. In contrast to Fig.~3 in the main text, we have artificially set $V_H=0$, to observe the isolated impact of the intrinsic curvature contribution of the Laplace-Beltrami operator, which is indeed subleading.

\section{Derivation of the tangential divergence}\label{supp:derivation_tangential_divergence}

\subsection{Field solutions}\label{supp:derivation_tangential_divergence_sec_1}
To get the tangential divergence $\nabla_a\mathcal{A}^a$ at the interface, we act with $\nabla_a$ from the left on Eq.~\eqref{eq:wave_equation_tangential}. Since $\epsilon(\eta)$ and $\partial_\eta$ depend only on the normal coordinate, they commute with the tangential derivative $\nabla_a$, but not necessarily with $\Delta_{\gamma}$. The commutator is (see Section~\ref{supp:commutator_covariant_laplace})
\begin{equation}
    \left[\nabla_a, \Delta_{\gamma}\right]\,\mathcal{A}^a=(\nabla_a K) \mathcal{A}^a + K (\nabla_a \mathcal{A}^a),
\end{equation}
which we can drop as it is $\mathcal{O}(K)=\mathcal{O}(\alpha^2)$. We also drop $\nabla_aC^a$, see Section~\ref{supp:wave_equation_curved_space}. We then get
\begin{equation}\label{eq:nabla_a_on_tangential_equation}
    \begin{aligned}
    \nabla_a\left[\Delta_\gamma + 2\eta\mathcal{D}_h + \partial_\eta^2 + k_0^2\epsilon\right]\mathcal{A}^a - 2\nabla_a\left( h^a_d \partial_\eta \mathcal{A}^d\right) &=2 \nabla_a\left(h^{ab} \partial_b \mathcal{A}^\eta \right),\\
        \left[\Delta_\gamma + \partial_\eta^2 + k_0^2\epsilon\right]\nabla_a\mathcal{A}^a  +2\eta \nabla_a\mathcal{D}_h\mathcal{A}^a - 2 h^a_d \nabla_a\left(\partial_\eta \mathcal{A}^d\right) &\approx 2h^{ab} \nabla_a \nabla_b \mathcal{A}^\eta=2\mathcal{D}_h\mathcal{A}^\eta.
    \end{aligned}
\end{equation}
The second and third term in the last equation are at least $\mathcal{O}(\alpha)$ as
\begin{equation}
    2\eta \nabla_a\mathcal{D}_h\mathcal{A}^a - 2 h^a_d \nabla_a\left(\partial_\eta \mathcal{A}^d\right)=2\alpha\left[ \eta \nabla_a\mathcal{D}_h^{(1)}\mathcal{A}^{a\,(0)} - h^{a\,(1)}_d \nabla_a\left(\partial_\eta \mathcal{A}^{d\,(0)}\right)\right]+\mathcal{O}(\alpha^2).
\end{equation}
As the flat SPP mode has no tangential components, i.e., $\mathcal{A}^{a\,(0)}=\mathcal{A}^{d\,(0)}=0$, see Section~\ref{sec:flat_spp_and_gauge}, this sum does not further contribute in our derivation. We now define $\Xi\equiv \nabla_a\mathcal{A}^a$ and Eq.~\eqref{eq:nabla_a_on_tangential_equation} defines an ordinary differential equation (ODE) for $\Xi$ up to $\mathcal{O}(\alpha)$ as
\begin{equation}
    \hat{O}^{(0)}\,\Xi =2\alpha\mathcal{D}_h^{(1)}\,\mathcal{A}^{\eta\,(0)},
\end{equation}
with
\begin{equation}
    \hat{O}^{(0)}\equiv \partial_\eta^2 + k_0^2 \epsilon(\eta) + \Delta_{\rm flat} .
\end{equation}
Expanding also the yet unknown $\Xi$ in orders of $\alpha$, we write
\begin{equation}
    \hat{O}^{(0)} \left( \Xi^{(0)} + \alpha \Xi^{(1)} \right) 
    =
    2\alpha f(\eta) \mathcal{D}_h^{(1)} \psi^{(0)}
\end{equation}
for which we used $\mathcal{A}^{\eta}$ only up to zero order in $\alpha$ on the RHS and the shorthand notation $\mathcal{A}^{\eta\,(0)}=f(\eta)\psi^{(0)}(q)$, see Eq.~\eqref{eq:separation_Az}. At zero order, we get
\begin{equation}
    \hat{O}^{(0)} \Xi^{(0)} = 0, \quad\implies \Xi^{(0)}=0,
\end{equation} 
in line with the flat SPP physics, see Section~\ref{sec:flat_spp_and_gauge}. At first order in $\alpha$, the ODE then reads
\begin{equation}
    \hat{O}^{(0)}  \Xi^{(1)} 
    =
    \left[ \partial_\eta^2 + k_0^2 \epsilon(\eta) + \Delta_{\rm flat} \right] \Xi^{(1)} 
    =
    2 f(\eta)\mathcal{D}_h^{(1)} \psi^{(0)}.
\end{equation}
Let us define the tangential source function as
\begin{equation}
    S^{(1)}(q_\parallel) \equiv 2\mathcal{D}_h^{(1)}\psi^{(0)}= 2\left(H^{(1)}\Delta_{\rm flat} + \sigma^{ab\,(1)}\partial_a\partial_b\right) \psi^{(0)},
\end{equation}
to keep the notation compact. Because the unperturbed SPP envelope satisfies the flat wave equation $\Delta_{\rm flat} \psi^{(0)} = -k_{\rm spp}^2 \psi^{(0)}$, and because $\Delta_{\rm flat}$ commutes with the constant curvature coefficients and the partial derivatives $\partial_a\partial_b$, the source function $S^{(1)}(q_\parallel)$ is itself an eigenfunction of the flat Laplacian with the same eigenvalue, i.e., 
\begin{equation}
    \Delta_{\rm flat} S^{(1)}(q_\parallel) = -k_{\rm spp}^2 S^{(1)}(q_\parallel).
\end{equation}
By making the separable ansatz $\Xi^{(1)}(q_\parallel, \eta) = \chi(\eta) S^{(1)}(q_\parallel)$, we therefore arrive at
\begin{equation}
    \left[ \partial_\eta^2 + k_0^2 \epsilon(\eta) - k_{\rm spp}^2 \right] \chi(\eta) S^{(1)}(q_\parallel) = f(\eta) S^{(1)}(q_\parallel),
\end{equation}
or, using $\kappa_i^2 = k_{\rm spp}^2 - k_0^2 \epsilon_i$ from Eq.~\eqref{eq:kappa_wavenumber}, at the differential equation for the vertical component $\eta$ as
\begin{equation}
    \left[ \partial_\eta^2 - \kappa_i^2 \right] \chi_i(\eta) = f_i(\eta),
\end{equation}
within each domain. Solving this by standard techniques and requiring continuity of $\chi$ at the interface yields
\begin{equation}
    \begin{aligned}
    \Xi_d^{(1)}(q_\parallel, \eta) = \left[ C - \frac{\eta}{2\kappa_d}  \right] e^{-\kappa_d \eta} S^{(1)}(q_\parallel) \quad \text{for } \eta > 0,
    \\ 
    \Xi_m^{(1)}(q_\parallel, \eta) = \left[ C + \frac{\eta}{2\kappa_m}  \right] e^{\kappa_m \eta} S^{(1)}(q_\parallel)  \quad \text{for } \eta < 0.
\end{aligned}
\end{equation}
Continuity of $\partial_\eta\chi_i$ at the interface yields
\begin{equation}
    C=-\frac{1}{2\kappa_d\kappa_m}=-\frac{1}{2k^2_{\rm spp}}.
\end{equation}
Hence,
\begin{equation}
    \Xi^{(1)}(q_\parallel, \eta) 
    =\nabla_a\mathcal{A}^a(q_\parallel, \eta)=
    \left(-H^{(1)}k^2_{\rm spp} + \sigma^{ab\,(1)}\partial_a\partial_b\right) \psi^{(0)}
    \begin{cases} 
    \left( -\frac{1}{k^2_{\rm spp}} - \frac{\eta}{\kappa_d}\right)e^{-\kappa_d \eta} & \text{for } \eta > 0,\\
    \left( -\frac{1}{k^2_{\rm spp}} +\frac{\eta}{\kappa_m} \right)e^{\kappa_m \eta} & \text{for } \eta < 0,
    \end{cases}
\end{equation}
and at the interface $\eta=0$, we get
\begin{equation}
    \Xi^{(1)}(q_\parallel, 0) 
    =
    -\frac{1}{k^2_{\rm spp}}\left(-H^{(1)}k^2_{\rm spp} + \sigma^{ab\,(1)}\partial_a\partial_b\right) \psi^{(0)}
    =
    H^{(1)}\psi^{(0)}-\frac{1}{k^2_{\rm spp}}\sigma^{ab\,(1)}\partial_a\partial_b\psi^{(0)}.
\end{equation}

\subsection{Commuting covariant derivative and Laplace-Beltrami operator}\label{supp:commutator_covariant_laplace}
We want to calculate the difference 
\begin{equation}\label{eq:commutator}
   \left [\nabla_a, \Delta_{\gamma}\right] \mathcal{A}^a \equiv \nabla_a \left(\Delta_{\gamma} \mathcal{A}^a\right) - \Delta_{\gamma} \left(\nabla_a \mathcal{A}^a\right),
\end{equation}
with $\Delta_{\gamma} = \nabla_b \nabla^b$. We expand the first term as
\begin{equation}
    \nabla_a \left(\nabla_b \nabla^b \mathcal{A}^a\right)\equiv \nabla_a \left(\nabla_b T^{ba}\right).
\end{equation}
In the following, using the definition of the Riemann curvature tensor as in Ref.~\cite{misner1973gravitation}, we need the Ricci identities 
\begin{equation}\label{eq:id_1}
[\nabla_a, \nabla_b]\,T^{ba} = R^b_{c a b} T^{ca} + R^a_{c a b} T^{bc}, 
\end{equation}
and
\begin{equation}
    [\nabla_a, \nabla_b]\,\mathcal{A}^c=R^c_{dab}\,\mathcal{A}^d.
\end{equation}
Moreover, we need to use that for a 2D surface, dictated by the Bianchi identities, the Riemann tensor simplifies to
\begin{equation}\label{eq:id_2}
    R^a_{bcd}=K(\delta^a_c\gamma_{bd}-\delta^a_d\gamma_{bc}).
\end{equation}
This leads to
\begin{equation}
[\nabla_a, \nabla_b]\,T^{ba} = -K\gamma_{ca} T^{ca} + K\gamma_{cb} T^{bc}=K \left(T_c^{c}-T_a^{a}\right)=0. 
\end{equation}
Therefore,
\begin{equation}
    \nabla_a \nabla_b \,T^{ba}=\nabla_b \nabla_a \,T^{ba}, 
\end{equation}
and hence,
\begin{equation}
    \nabla_a \nabla_b \nabla^b \mathcal{A}^a=\nabla_b \nabla_a \nabla^b \mathcal{A}^a.
\end{equation}
Now we need
\begin{equation}
    [\nabla_a, \nabla^b] \mathcal{A}^a = \gamma^{bc} [\nabla_a, \nabla_c] \mathcal{A}^a = \gamma^{bc} R^a_{d a c} \mathcal{A}^d,
\end{equation}
which after a short calculation using Eq.~\eqref{eq:id_2} gives us
\begin{equation}
    [\nabla_a, \nabla^b]\,\mathcal{A}^a=K\,\mathcal{A}^b.
\end{equation}
Therefore, the first term in Eq.~\eqref{eq:commutator} yields
\begin{equation}
    \begin{aligned}
    \nabla_a \left(\Delta_{\gamma} \mathcal{A}^a\right)&=\nabla_b \nabla_a \nabla^b \mathcal{A}^a\\
    &=\nabla_b\left( \nabla^b \nabla_a \mathcal{A}^a + K \mathcal{A}^b\right)\\
    &=\Delta_{\gamma} (\nabla_a \mathcal{A}^a) + (\nabla_b K) \mathcal{A}^b + K (\nabla_b \mathcal{A}^b).
\end{aligned}
\end{equation}
Eq.~\eqref{eq:commutator} therefore becomes
\begin{equation}
     \left [\nabla_a, \Delta_{\gamma}\right] \mathcal{A}^a=(\nabla_b K) \mathcal{A}^b + K (\nabla_b \mathcal{A}^b).
\end{equation}

\section{Surface Green's function and collective observables}\label{supp:collective_observables}
For emitters at distances of a few tens of nanometers from a flat metal surface, the SPP channel dominates the dipole-dipole interaction over both the radiative and the lossy contributions, cf. Eq.~(2.42) in Ref.~\cite{chance1978molecular} and Fig.~8(a) in Ref.~\cite{barnes1998fluorescence}, which both treat fluorescence near interfaces. The coherent and dissipative rates entering Eq.~(6) of the main text are determined by the dyadic Green's tensor $G$ via~\cite{asenjo2017exponential}
\begin{equation}\label{eq:interaction_rates}
    J^{ij} - i\frac{\Gamma^{ij}}{2} = -\frac{\mu_0\omega_0^2\wp^2}{\hbar}\,G_{\eta\eta}(\mathbf{r}_{0,i},\mathbf{r}_{0,j},\omega_0),
\end{equation}
with $\wp$ the transition dipole moment. Since the dipole moments are oriented normally to the metal surface, the interatomic coupling is governed by the normal-normal component $G_{\eta\eta}$ evaluated at the respective emitter positions.

On a flat interface, the SPP contribution to $G_{\eta\eta}=G_{zz}$ follows by decomposing the free-space dyadic Green's tensor into plane waves via the Weyl identity~\cite{novotny2012principles, chance1978molecular} and extracting the residue at the SPP pole of the TM Fresnel reflection coefficient, which yields the factorized form
\begin{equation}\label{eq:factorisation_G_flat}
    G_{zz} = C_0\, e^{-2\kappa_d d}\, G^{\rm flat}_{\rm 2D},
\end{equation}
with the flat-surface Green's function
\begin{equation}\label{eq:G_flat_2D}
    G^{\rm flat}_{\rm 2D}(\mathbf{q},\mathbf{q}_0) = \frac{i}{4}H^{(1)}_0\!\left(k_{\rm spp}|\mathbf{q}-\mathbf{q}_0|\right),
\end{equation}
$H^{(1)}_0$ the Hankel function of the first kind, and $C_0$ a geometry-independent material constant derived in Section~\ref{supp:derivation_constant_C0}. Equation~\eqref{eq:factorisation_G_flat} reflects the separability $\mathcal{A}^z=f(z)\psi(x,y)$ of the flat-interface mode, cf. Eq.~\eqref{eq:separation_Az}.

On a curved interface, the normal-normal component $G_{\eta\eta}$, does not separate exactly, owing to the polynomial correction terms. However, for emitter-surface distances $d$ of a few to several tens of nanometers, as realized for dye molecules, quantum dots, and nitrogen-vacancy centers~\cite{wei2011quantum, rose2014control, li2015quantum, huck2011controlled}, the condition $\kappa_d d \ll 1$ holds. In this limit the terms $A_d d^2$ and $B_d d$ in Eq.~\eqref{eq:ansatz_A_eta} vanish, and the remaining $C_d\psi^{(0)}$ is a multiplicative constant that can be absorbed, leaving the Green's function structure unchanged. Equation~\eqref{eq:factorisation_G_flat} then carries over as
\begin{equation}\label{eq:factorisation_G_curved}
    G_{\eta\eta} \approx C_0\, G^{\rm curved}_{\rm 2D}(\mathbf{q}_i,\mathbf{q}_j),
\end{equation}
with $G^{\rm curved}_{\rm 2D}$ defined by Eq.~(7) of the main text.

The diagonal elements ($i=j$) define the single-emitter decay rate into the SPP channel. Since $G^{\rm curved}_{\rm 2D}(\mathbf{q}_0,\mathbf{q}_0)$ has a logarithmically divergent real part, a standard point-dipole artifact that would be regularized by any finite emitter size, the self-energy contribution $\mathrm{Re}[C_0]\,\mathrm{Re}[G^{\rm curved, self}_{\rm 2D}]$ to the frequency shift $J^{ii}$ is absorbed into the physically measured transition frequency $\omega_0$ by renormalization~\cite{asenjo2017exponential}. For a lossy metal with complex $C_0$, the product $\mathrm{Im}[C_0]\,\mathrm{Re}[G^{\rm curved, self}_{\rm 2D}]$ produces an analogous divergent contribution to the decay rate, which is likewise absorbed into the observed single-emitter rate. The renormalized single-emitter decay rate is therefore defined as
\begin{equation}\label{eq:gamma_0_curved}
    \gamma_0^{\rm curved} \equiv \Gamma^{ii} = \frac{2\mu_0\omega_0^2\wp^2}{\hbar}\,\mathrm{Re}[C_0]\,\mathrm{Im}\!\left[G^{\rm curved, self}_{\rm 2D}\right].
\end{equation}
For the off-diagonal elements ($i\neq j$), both the real and imaginary parts of $G^{\rm curved}_{\rm 2D}$ are finite, and the contribution from $\mathrm{Im}[C_0]$ can additionally be neglected because $\mathrm{Im}[C_0]/\mathrm{Re}[C_0]\lesssim 5\,\%$ for typical noble metals at optical frequencies~\cite{johnson1972optical}, as shown at the end of Section~\ref{supp:derivation_constant_C0}.

Writing the $k$-th eigenvalue of the interaction matrix as $\lambda_k = \Delta_k - i\gamma_k/2$, the collective decay rates and cooperative frequency shifts are $\gamma_k = -2\,\mathrm{Im}[\lambda_k]$ and $\Delta_k = \mathrm{Re}[\lambda_k] - J^{ii}$. Since the self-interaction $J^{ii}$ is identical for each collective mode $k$, subtracting it isolates the mode-dependent cooperative shift, and the normalized observables read
\begin{equation}\label{eq:normalized_rates}
    \frac{\gamma_k}{\gamma_0^{\rm curved}} = \frac{-2\,\mathrm{Im}[\lambda_k]}{\gamma_0^{\rm curved}},
    \qquad
    \frac{\Delta_k}{\gamma_0^{\rm curved}} = \frac{\mathrm{Re}[\lambda_k]-J^{ii}}{\gamma_0^{\rm curved}}.
\end{equation}
Normalizing to $\gamma_0^{\rm curved}$ on the same surface also removes the weak curvature dependence of the single-emitter rate itself, which would otherwise enter for larger emitter distances.

\section{SPP-mediated emitter-emitter interaction on a metal spheroid}\label{supp:greens_interaction_full}

\subsection{Wave function on the spheroid surface}
We parametrize the spheroid using angular coordinates $(\theta,\phi)$, with $\theta\in[0,\pi]$ the polar angle measured from the north pole and $\phi\in[0,2\pi)$ the azimuthal angle. The equatorial and polar semi-axes are $a$ and $c$, respectively, with $a=c=R$ recovering the sphere. The embedding is
\begin{equation}
\mathbf{r}(\theta,\phi)=
\begin{pmatrix}
a\sin\theta\cos\phi\\
a\sin\theta\sin\phi\\
c\cos\theta
\end{pmatrix}.
\end{equation}
The tangent vectors $\mathbf{e}_a=\partial_a\mathbf{r}$ give the first fundamental form $\gamma_{ab}=\mathbf{e}_a\cdot\mathbf{e}_b$,
\begin{equation}
\gamma_{ab}=
\begin{pmatrix}
\gamma_{\theta\theta} & 0\\
0 & \gamma_{\phi\phi}
\end{pmatrix}
=
\begin{pmatrix}
\rho & 0\\
0 & a^2\sin^2\theta
\end{pmatrix},
\qquad
\rho\equiv a^2\cos^2\theta+c^2\sin^2\theta,
\end{equation}
with $\sqrt{\gamma}=a\sin\theta\sqrt{\rho}$ and inverse metric $\gamma^{\theta\theta}=1/\rho$, $\gamma^{\phi\phi}=1/(a^2\sin^2\theta)$. The first fundamental form and all quantities derived from it alone are intrinsic and independent of the choice of normal orientation. To define the second fundamental form we must choose the unit normal $\hat{\mathbf{n}}$. We introduce the sign parameter
\begin{equation}
s = \begin{cases} 
+1 & \hat{\mathbf{n}}\text{\, points outwards = convex metal surface},\\ -1 & \hat{\mathbf{n}}\text{\, points inwards = concave metal surface},
\end{cases}
\end{equation}
and write
\begin{equation}
\hat{\mathbf{n}} = s\,\frac{\mathbf{e}_\theta\times\mathbf{e}_\phi}{|\mathbf{e}_\theta\times\mathbf{e}_\phi|} = s\,\frac{\mathbf{e}_\theta\times\mathbf{e}_\phi}{\sqrt{\gamma}}.
\end{equation}
The outward cross product evaluates to
\begin{equation}
\mathbf{e}_\theta\times\mathbf{e}_\phi =
\begin{pmatrix}
ac\sin^2\theta\cos\phi\\
ac\sin^2\theta\sin\phi\\
a^2\sin\theta\cos\theta
\end{pmatrix},
\end{equation}
so that $|\mathbf{e}_\theta\times\mathbf{e}_\phi|=a\sin\theta\sqrt{\rho}=\sqrt{\gamma}$, which yields
\begin{equation}
    \hat{\mathbf{n}} = \frac{s}{\sqrt{\rho}} \begin{pmatrix} c \sin\theta \cos\phi \\ c \sin\theta \sin\phi \\ a \cos\theta \end{pmatrix}.
\end{equation}
The second fundamental form $h_{ab}=\hat{\mathbf{n}}\cdot\partial_a\partial_b\mathbf{r}$ is calculated to
\begin{equation}
h_{ab}=
\begin{pmatrix}
h_{\theta\theta} & 0\\
0 & h_{\phi\phi}
\end{pmatrix}
= -\frac{s}{\sqrt{\rho}}
\begin{pmatrix}
ac & 0\\
0 & ac\sin^2\theta
\end{pmatrix} \implies 
h^{ab} = -s\begin{pmatrix} \frac{ a c}{\rho^{5/2}} & 0 \\ 0 & \frac{ c}{a^3 \sin^2\theta \sqrt{\rho}} \end{pmatrix}.
\end{equation}
The extrinsic curvature is then
\begin{equation}\label{eq:H_spheroid}
    H = \frac{1}{2} \gamma^{ab} h_{ab}=\frac{1}{2}\left(\gamma^{\theta\theta} h_{\theta\theta}+\gamma^{\phi\phi} h_{\phi\phi} \right)=-\frac{s}{2}\frac{c(a^2+\rho)}{a\,\rho^{3/2}},
\end{equation}
which is negative for a convex surface ($s=+1$) and positive for a concave surface ($s=-1$), and becomes $H=-s/R$ for the sphere at $a=c=R$. The traceless part of the shape operator, $\sigma^{ab}=h^{ab}-H\gamma^{ab}$, evaluates to
\begin{equation}
\sigma^{\theta\theta}=\frac{sc(c^2-a^2)\sin^2\theta}{2a\,\rho^{5/2}},\qquad \sigma^{\phi\phi}=\frac{sc(a^2-c^2)}{2a^3\rho^{3/2}},\qquad\sigma^{\theta\phi}=\sigma^{\phi\theta}=0.
\end{equation}
We further need the Christoffel symbols for the covariant operator $\sigma^{ab}\nabla_a\nabla_b$. Expanding $\nabla_a \nabla_b \psi = \partial_a \partial_b \psi - \Gamma^c_{ab} \partial_c \psi$ for the spheroid yields
\begin{equation}
    \begin{aligned}
    \sigma^{ab} \nabla_a \nabla_b \psi 
    &= \sigma^{\theta\theta} \nabla_\theta \nabla_\theta \psi + \sigma^{\phi\phi} \nabla_\phi \nabla_\phi \psi\\
    &= \sigma^{\theta\theta} \left(\partial_\theta^2 \psi - \Gamma^\theta_{\theta\theta} \partial_\theta \psi - \Gamma^\phi_{\theta\theta} \partial_\phi \psi \right)+ \sigma^{\phi\phi} \left(\partial_\phi^2 \psi - \Gamma^\theta_{\phi\phi} \partial_\theta \psi - \Gamma^\phi_{\phi\phi} \partial_\phi \psi \right).
\end{aligned}
\end{equation}
Using 
\begin{equation}
    \Gamma^a_{bc} = \frac{1}{2}\gamma^{ad} (\partial_b \gamma_{dc} + \partial_c \gamma_{bd} - \partial_d \gamma_{bc}),
\end{equation}
and
\begin{equation}
    \begin{aligned}
    &\partial_\theta \gamma_{\theta\theta} = \partial_\theta (a^2\cos^2\theta + c^2\sin^2\theta)=2(c^2 - a^2)\sin\theta\cos\theta,\\
    &\partial_\theta \gamma_{\phi\phi} = \partial_\theta (a^2\sin^2\theta) = 2a^2\sin\theta\cos\theta,
\end{aligned}
\end{equation}
the required Christoffel symbols are calculated as
\begin{equation}
    \Gamma^\theta_{\theta\theta} = \frac{(c^2 - a^2)\sin\theta\cos\theta}{\rho},\qquad 
    \Gamma^\phi_{\theta\theta} = 0,\qquad
    \Gamma^\theta_{\phi\phi} = -\frac{a^2\sin\theta\cos\theta}{\rho},\qquad 
    \Gamma^\phi_{\phi\phi}= 0,
\end{equation}
so that
\begin{equation}
    \sigma^{ab} \nabla_a \nabla_b \psi=\sigma^{\theta\theta} \left(\partial_\theta^2 \psi - \Gamma^\theta_{\theta\theta} \partial_\theta \psi \right)+ \sigma^{\phi\phi} \left(\partial_\phi^2 \psi - \Gamma^\theta_{\phi\phi} \partial_\theta \psi\right)=\sigma^{\theta\theta}\partial_\theta^2 \psi+\sigma^{\phi\phi} \partial_\phi^2 \psi -\left( \sigma^{\theta\theta}\Gamma^\theta_{\theta\theta} +\sigma^{\phi\phi}\Gamma^\theta_{\phi\phi}\right) \partial_\theta \psi,
\end{equation}
with
\begin{equation}
    \sigma^{\theta\theta}\Gamma^\theta_{\theta\theta}=\frac{sc(c^2-a^2)^2\sin^3\theta \cos\theta}{2a\,\rho^{7/2}},\qquad \sigma^{\phi\phi}\Gamma^\theta_{\phi\phi}=\frac{sc(c^2-a^2)\sin\theta\cos\theta}{2a\rho^{5/2}}.
\end{equation}
The full anisotropic operator (Eq.~(4b) in the main text) is then
\begin{equation}
C_\sigma\sigma^{ab}\nabla_a\nabla_b = sC_\sigma\left[\frac{c(c^2-a^2)\sin^2\theta}{2a\,\rho^{5/2}}\partial_\theta^2 + \frac{c(a^2-c^2)}{2a^3\rho^{3/2}}\partial_\phi^2 - \frac{c(c^2-a^2)\sin\theta\cos\theta}{2a\rho^{5/2}}\left(1+\frac{(c^2-a^2)\sin^2\theta}{\rho} \right)\partial_\theta \right].
\end{equation}
The Laplace-Beltrami operator, being intrinsic, is independent of $s$, and reads
\begin{equation}
\Delta_\gamma=\frac{1}{\sqrt{\gamma}}\partial_a\left(\sqrt{\gamma}\,\gamma^{ab}\partial_b\right)=\frac{1}{\rho}\partial_\theta^2+\frac{a^2\cot\theta}{\rho^2}\partial_\theta+\frac{1}{a^2\sin^2\theta}\partial_\phi^2.
\end{equation}
The curvature potential (Eq.(4a) in the main text) becomes
\begin{equation}
V_{\rm H} = C_H H = -\frac{sC_H c(a^2+\rho)}{2a\,\rho^{3/2}}.
\end{equation}
Collecting all terms, the wave equation on the spheroid reads
\begin{equation}
    \begin{aligned}\label{eq:wave_spheroid}
\Bigg\{&\left[\frac{1}{\rho}
+sC_\sigma\frac{c(c^2-a^2)\sin^2\theta}{2a\,\rho^{5/2}}\right]\partial_\theta^2
+\left[\frac{1}{a^2\sin^2\theta}
+sC_\sigma\frac{c(a^2-c^2)}{2a^3\rho^{3/2}}\right]\partial_\phi^2 \\
&+\left[\frac{a^2\cot\theta}{\rho^2}- sC_\sigma\frac{c(c^2-a^2)\sin\theta\cos\theta}{2a\rho^{5/2}}\left(1+\frac{(c^2-a^2)\sin^2\theta}{\rho} \right)\right]\partial_\theta +k_{\rm spp}^2-sC_H\frac{c(a^2+\rho)}{2a\,\rho^{3/2}}\Bigg\}\psi=0.
\end{aligned}
\end{equation}
Note that $C_H$ and $C_\sigma$ are complex numbers for a lossy metal with $\mathrm{Im}(\epsilon_m)\neq 0$. In general, the Green's function~$G(\theta,\phi;\theta_0,\phi_0)$ satisfies
\begin{equation}\label{eq:operator_equation_L_G}
\hat{L}\,G(\theta,\phi;\theta_0,\phi_0)=-\frac{\delta(\theta-\theta_0)\delta(\phi-\phi_0)}{a\sin\theta\sqrt{\rho}}
\end{equation}
with the RHS representing a single point source located at $(\theta_0, \phi_0)$. Since all coefficients of~$\hat{L}$ are independent of~$\phi$, as a result of the rotational symmetry, the azimuthal eigenfunctions are plane waves~$e^{im\phi}$. Therefore, we can expand
\begin{equation}\label{eq:greens_function_fourier}
G(\theta,\phi;\theta_0,\phi_0)=\frac{1}{2\pi}\sum_{m=-\infty}^{\infty}g_m(\theta,\theta_0)\,e^{im(\phi-\phi_0)},
\end{equation}
using the completeness relation $\delta(\phi-\phi_0)=\frac{1}{2\pi}\sum_{m}e^{im(\phi-\phi_0)}$ for the RHS in Eq.~\eqref{eq:operator_equation_L_G}. We thus get
\begin{equation}
    \frac{1}{2\pi} \sum_{m=-\infty}^{\infty} \hat{L} \Big( g_m(\theta, \theta_0) e^{im(\phi-\phi_0)} \Big) = \frac{1}{2\pi} \sum_{m=-\infty}^{\infty} \left[ -\frac{\delta(\theta-\theta_0)}{a\sin\theta\sqrt{\rho}} \right] e^{im(\phi-\phi_0)}.
\end{equation}
$\hat{L}$ is the differential operator in Eq.~\eqref{eq:wave_spheroid}. Defining $\hat{L}_m = A(\theta)\partial_\theta^2 + C(\theta)\partial_\theta + C_m(\theta)$, we get
\begin{equation}
    \sum_{m=-\infty}^{\infty} \left\{ \hat{L}_m g_m(\theta, \theta_0) + \frac{\delta(\theta-\theta_0)}{a\sin\theta\sqrt{\rho}} \right\} e^{im(\phi-\phi_0)} = 0.
\end{equation}
Since the functions $e^{im(\phi-\phi_0)}$ form a complete orthogonal basis, we require
\begin{equation}
    \hat{L}_m g_m(\theta, \theta_0) = -\frac{\delta(\theta-\theta_0)}{a\sin\theta\sqrt{\rho}},
\end{equation}
for each m individually. Writing everything out, yields the one-dimensional ODE
\begin{equation}
    \begin{aligned}\label{eq:radial_ode}
&\Bigg\{\left[\frac{1}{\rho}+s\,C_\sigma\frac{c(c^2-a^2)\sin^2\theta}{2a\,\rho^{5/2}}\right]\partial_\theta^2
+\left[\frac{a^2\cot\theta}{\rho^2}- sC_\sigma\frac{c(c^2-a^2)\sin\theta\cos\theta}{2a\rho^{5/2}}\left(1+\frac{(c^2-a^2)\sin^2\theta}{\rho} \right)\right]\partial_\theta\\
&-m^2\left[\frac{1}{a^2\sin^2\theta}+s\,C_\sigma\frac{c(a^2-c^2)}{2a^3\rho^{3/2}}\right]+k_{\rm spp}^2-s\,C_H\frac{c(a^2+\rho)}{2a\,\rho^{3/2}}\Bigg\}g_m(\theta,\theta_0)=-\frac{\delta(\theta-\theta_0)}{a\sin\theta\sqrt{\rho}}.
\end{aligned}
\end{equation}
Note that the $m$-dependence of the operator in Eq.~\eqref{eq:radial_ode} enters only as~$m^2$. The ODEs for modes~$+m$ and~$-m$ are therefore identical, which means $g_m=g_{-m}$ for all~$m\in\mathbb{Z}$.

\subsection{Sommerfeld radiation condition and perfectly matched layer}
This section provides the preparation for the numerical computation. To enforce the Sommerfeld radiation condition~\cite{sommerfeld1949partial}, meaning that waves propagate outward from the source without reflection, we introduce a perfectly matched layer~(PML)~\cite{berenger1994perfectly} in the region~$\theta\in(\theta_{\rm PML},\theta_{\rm max}]$ with $\theta_{\rm PML}>\theta_0$ through a complex coordinate stretching. For the PML, we define the stretched coordinate as
\begin{equation}
d\tilde{\theta}=\zeta(\theta)\,d\theta,\qquad
\zeta(\theta)=1+i\sigma(\theta),\qquad
\zeta'(\theta)=i\sigma'(\theta),\qquad
\sigma(\theta)=
\begin{cases}
0&\theta\leq\theta_{\rm PML},\\\sigma_{\rm max}\!\left(\dfrac{\theta-\theta_{\rm PML}}{\theta_{\rm max}-\theta_{\rm PML}}\right)^3&\theta>\theta_{\rm PML},\end{cases}
\end{equation}
with $\sigma(\theta)$ a smooth ramp-up of absorption and ~$\sigma_{\rm max}>0$ controls the absorption strength. The derivative operators transform as
\begin{equation}
\partial_{\tilde{\theta}}=\frac{1}{\zeta(\theta)}\partial_\theta,\qquad\partial_{\tilde{\theta}}^2=\frac{1}{\zeta^2(\theta)}\partial_\theta^2-\frac{\zeta'(\theta)}{\zeta^3(\theta)}\partial_\theta.
\end{equation}
We define the PML-modified coefficients
\begin{equation}
    \begin{aligned}
&A(\tilde{\theta})=\frac{1}{\tilde{\rho}}+s\,C_\sigma\frac{c(c^2-a^2)\sin^2\tilde{\theta}}{2a\,\tilde{\rho}^{5/2}},\\
&B(\tilde{\theta})=\frac{a^2\cot\tilde{\theta}}{\tilde{\rho}^2}- sC_\sigma\frac{c(c^2-a^2)\sin\tilde{\theta}\cos\tilde{\theta}}{2a\tilde{\rho}^{5/2}}\left(1+\frac{(c^2-a^2)\sin^2\tilde{\theta}}{\tilde{\rho}} \right),\\ 
&C_m(\tilde{\theta})=-m^2\!\left[\frac{1}{a^2\sin^2\tilde{\theta}}+s\,C_\sigma\frac{c(a^2-c^2)}{2a^3\tilde{\rho}^{3/2}}\right]+k_{\rm spp}^2-s\,C_H\frac{c(a^2+\tilde{\rho})}{2a\,\tilde{\rho}^{3/2}},
\end{aligned}
\end{equation}
with $\tilde\rho=a^2\cos^2\tilde\theta+c^2\sin^2\tilde\theta$. Eq.~\eqref{eq:radial_ode} is then rewritten as
\begin{equation}\label{eq:pml_ode}
\frac{A(\tilde{\theta})}{\zeta^2(\theta)}\,\partial_\theta^2\,g_m+\left[\frac{B(\tilde{\theta})}{\zeta(\theta)}-\frac{A(\tilde{\theta})\,\zeta'(\theta)}{\zeta^3(\theta)}\right]\partial_\theta\,g_m+C_m(\tilde{\theta})\,g_m=-\frac{\delta(\theta-\theta_0)}{a\sin\theta\sqrt{\rho}},
\end{equation}
with
\begin{equation}
    \tilde{\theta}=\tilde{\theta}(\theta)=\theta+i\int_0^\theta\sigma(\vartheta)d\vartheta,
\end{equation}
and with the source term of the emitter evaluated at the real coordinate~$\theta$, as the dipole sits outside the PML at $\theta_0<\theta_{\rm PML}$. Two additional boundary conditions must be introduced. For this, we examine the behavior of the ODE coefficients as $\theta\to 0$. Using $\sin\theta\approx\theta$, $\cos\theta\approx 1$ and $\rho(0)=a^2$, the two singular terms are 
\begin{equation} 
B(\theta) \approx \frac{1}{a^2\theta}, \qquad -\frac{m^2}{a^2\sin^2\theta}\approx-\frac{m^2}{a^2\theta^2}, 
\end{equation} 
so $\theta=0$ is a regular singular point of the ODE for all $m$. To handle this, we impose a Dirichlet boundary condition 
\begin{equation} 
g_m(0,\theta_0)=0, \qquad m\neq 0. 
\end{equation} 
For $m=0$, we impose a Neumann boundary condition as
\begin{equation} 
\partial_\theta g_0(\theta,\theta_0)\bigg|_{\theta=0}=0. 
\end{equation} 
While $g_m(\theta)$ is continuous at $\theta=\theta_0$, there is a jump for the derivative, which we obtain by integrating Eq.~\eqref{eq:pml_ode} over $[\theta_0-\varepsilon,\theta_0+\varepsilon]$ and then let $\varepsilon\to 0$. The terms involving $g_m$ and $\partial_\theta g_m$ vanish in this limit and only the $\partial_\theta^2$ term survives, yielding 
\begin{equation} 
A(\theta_0)\Bigl[\partial_\theta g_m\Bigr]_{\theta_0^-}^{\theta_0^+}=-\frac{1}{a\sin\theta_0\sqrt{\rho_0}}.
\end{equation} 
We discretize the continuous domain $\theta \in [0, \theta_{\rm max}]$ using second-order central finite differences, which maps the differential operator to a tridiagonal matrix. The Neumann boundary condition for $m=0$ is enforced up to second-order accuracy using a ghost-node method to prevent artificial back scattering at the pole. Finally, the source term, which is responsible for the jump in $\partial_\theta g_m$, is projected onto the discrete grid using a linear interpolation weighting. The full discrete Green's function $g_m(\theta, \theta_0)$ is then obtained by a numerical inversion of the resulting linear system.

\subsection{Collective eigenvalues (Recap)} 
We place the $N$ emitters in a ring configuration at a fixed polar angle $\theta_0$. As they are equally spaced along the azimuthal direction, the surface coordinates of the $j$-th emitter are $\mathbf{q}_0 = (\theta_0, \phi_{0,j})$, with the azimuthal angles given by
\begin{equation}
    \phi_{0,j} = \frac{2\pi (j-1)}{N}, \qquad j = 1, 2, \ldots, N.
\end{equation}
The coupling between any two emitters in the ring is therefore written as
\begin{equation}
    \begin{aligned}
    \Omega^{jl} &=-\frac{\mu_0\omega_0^2\wp^2}{\hbar}G_{\eta\eta}(\mathbf{q}_j,\eta_j=d, \mathbf{q}_l,\eta_l=d ) \\
    &=-\frac{\mu_0\omega_0^2\wp^2}{\hbar}C_0\,e^{-2\kappa_d d}\,G_{\rm 2D}^{\rm curved}(\theta_0, \phi_{0,j}; \theta_0, \phi_{0,l}) = -\frac{\mu_0\omega_0^2\wp^2}{\hbar}C_0\,e^{-2\kappa_d d} \frac{1}{2\pi} \sum_{m=-\infty}^{\infty} g_m(\theta_0, \theta_0) e^{i m \frac{2\pi (j-l)}{N}},
\end{aligned}
\end{equation}
in which we inserted Eq.~\eqref{eq:greens_function_fourier} and used $e^{-2\kappa_d d}\approx 1$. The matrix defined by $\Omega^{jl}$ is therefore circulant. The eigenvectors are always the discrete Fourier basis vectors. Introducing the collective mode index $k \in \{0, 1, \ldots, N-1\}$, the $j$-th component of the $k$-th normalized eigenvector $\mathbf{v}^{(k)}$ is given by
\begin{equation}
    v_j^{(k)} = \frac{1}{\sqrt{N}} e^{i k \phi_{0,j}} = \frac{1}{\sqrt{N}} e^{i k \frac{2\pi (j-1)}{N}}.
\end{equation}
Because we know the eigenfunctions, we do not need to diagonalize the matrix $\Omega^{jl}$. Instead, from the eigenvalue equation, we get 
\begin{equation}
    \sum_{l=1}^N \Omega^{jl} v_l^{(k)} = \lambda_k v_j^{(k)}\qquad \implies \qquad \lambda_k = \sum_{l=1}^N \Omega^{1l} e^{i k \frac{2\pi (l-1)}{N}}.
\end{equation}
Substituting the expression for $\Omega^{1l}$ into the sum for $\lambda_k$ yields
\begin{equation}
    \begin{aligned}
    \lambda_k &= \sum_{l=1}^N \left[ -\frac{\mu_0\omega_0^2\wp^2}{\hbar} C_0 e^{-2\kappa_d d} \frac{1}{2\pi} \sum_{m=-\infty}^{\infty} g_m(\theta_0, \theta_0) e^{-i m \frac{2\pi (l-1)}{N}} \right] e^{i k \frac{2\pi (l-1)}{N}}\\
    &= -\frac{\mu_0\omega_0^2\wp^2}{\hbar} C_0 e^{-2\kappa_d d} \frac{1}{2\pi} \sum_{m=-\infty}^{\infty} g_m(\theta_0, \theta_0) \left[ \sum_{l=1}^N e^{i (k - m) \frac{2\pi (l-1)}{N}} \right].
\end{aligned}
\end{equation}
The sum over the index $l$ can also be written as
\begin{equation}
    \sum_{l=1}^N e^{i (k - m) \frac{2\pi (l-1)}{N}}=\sum_{l=1}^N \left( e^{i (k - m) \frac{2\pi}{N}} \right)^{l-1}= N \sum_{q=-\infty}^{\infty} \delta_{m, \, k - qN}.
\end{equation}
Therefore, we get
\begin{equation}
    \lambda_k = -\frac{\mu_0\omega_0^2\wp^2}{\hbar} C_0 e^{-2\kappa_d d} \frac{N}{2\pi} \sum_{q=-\infty}^{\infty} g_{k + qN}(\theta_0, \theta_0).
\end{equation}
At this point, we define
\begin{equation}
    S\equiv \sum_{m=-\infty}^{\infty} g_m(\theta_0, \theta_0), \qquad S_k\equiv \sum_{q=-\infty}^{\infty} g_{k + qN}(\theta_0, \theta_0),
\end{equation}
for easier notation. Both $\mathrm{Re}[S]$ and $\mathrm{Re}[S_k]$ diverge logarithmically, whereas $\mathrm{Im}[S]$ and $\mathrm{Im}[S_k]$ are finite. Let us write $\lambda_k = \Delta_k - i\frac{\gamma_k}{2}$. This means the collective decay rate of mode $k$ is
\begin{equation}
    \gamma_k = -2\,\mathrm{Im}[\lambda_k]
    = 
    \frac{\mu_0\omega_0^2\wp^2}{\pi\hbar} N \,\mathrm{Im}\left[ C_0 e^{-2\kappa_d d} S_k \right],
\end{equation}
in which we must keep $C_0$ and $e^{-2\kappa_d d}$ inside the $\mathrm{Im}[.]$, as they are complex when the metal permittivity $\epsilon_m$ is complex. In the main text, we pointed out that we need the curved self-Green's function for the normalization. Setting $\theta=\theta_0$ and $\phi=\phi_0$ in  Eq.~\eqref{eq:greens_function_fourier}, the single-emitter decay rate then reads
\begin{equation}
    \gamma_0^{\rm curved} \equiv \Gamma^{nn}=-2\mathrm{Im}\left[ \Omega^{nn}\right] =\frac{\mu_0 \omega_0^2 \wp^2}{\pi\hbar}\,\mathrm{Im}\left[ C_0 e^{-2\kappa_d d} S \right].
\end{equation}

As argued in the main text, we may use $e^{-2\kappa_d d}\approx 1$ and discard the divergent point-dipole artifact contributions $\mathrm{Im}\left[C_0\right]\,\mathrm{Re}\left[ S\right]$ and $\mathrm{Im}\left[C_0\right]\,\mathrm{Re}\left[ S_k\right]$. The  collective and single-emitter rates then evaluate to
\begin{equation}
    \gamma_k = \frac{\mu_0\omega_0^2\wp^2}{\pi\hbar} N \,\mathrm{Re}\left[C_0\right]\,\mathrm{Im}\left[S_k\right],
    \qquad
    \gamma_0^{\rm curved} = \frac{\mu_0 \omega_0^2 \wp^2}{\hbar\pi} \,\mathrm{Re}\left[C_0\right]\, \mathrm{Im}\!\left[S \right],
\end{equation}
respectively, yielding the normalized collective decay rate for mode $k$ as
\begin{equation}\label{eq:normalized_gamma_k}
    \frac{\gamma_k}{\gamma_0^{\rm curved}} = N \frac{\mathrm{Im}\!\left[ S_k \right]}{\mathrm{Im}\!\left[ S \right]}.
\end{equation}
For the cooperative frequency shift, we start from $\Delta_k = \mathrm{Re}[\lambda_k]$, which gives
\begin{equation}
    \Delta_k = -\frac{\mu_0\omega_0^2\wp^2}{2\pi\hbar}N \,\mathrm{Re}\!\left[ C_0 S_k \right].
\end{equation}
This shift includes the single-particle self-interaction
\begin{equation}
    J^{ii} \equiv \mathrm{Re}[\Omega^{ii}]=-\frac{\mu_0 \omega_0^2 \wp^2}{2\pi\hbar}\,\mathrm{Re}\left[ C_0 S \right],
\end{equation}
whose divergent part is already absorbed into $\omega_0$, as argued in the main text. Subtracting it gives the finite cooperative frequency shift
\begin{equation}
    \Delta_k^{\rm coop} = \Delta_k - J^{ii} = -\frac{\mu_0\omega_0^2\wp^2}{2\pi\hbar}  \,\mathrm{Re}\!\left[ C_0 \left( N S_k - S \right) \right]= -\frac{\mu_0\omega_0^2\wp^2}{2\pi\hbar}  
    \left\{ 
    \,\mathrm{Re}\!\left[ C_0 \right]\mathrm{Re}\!\left[NS_k-S \right]-\mathrm{Im}\!\left[ C_0 \right]\mathrm{Im}\!\left[NS_k-S \right]
    \right\}.
\end{equation}
Normalizing by $\gamma_0^{\rm curved}$ and using that $\mathrm{Im}\left[C_0\right]/\mathrm{Re}\left[C_0\right]\ll 1$ for typical noble metals, see Section~\ref{supp:derivation_constant_C0}, gives
\begin{equation}\label{eq:normalized_Delta_k}
    \frac{\Delta_k^{\rm coop}}{\gamma_0^{\rm curved}} = -\frac{1}{2} \frac{\mathrm{Re}\!\left[N S_k - S\right]}{\mathrm{Im}\!\left[ S \right]}.
\end{equation}

\subsection{Role of the linear potential}\label{supp:VH_zero_test}
To isolate the role of the linear-in-curvature potentials in Eq.~(1) of the main text, we repeat the calculation for the emitter ring on the sphere with $V_H$ artificially set to zero, so that curvature enters only through the intrinsic geometry contained in $\Delta_\gamma$. The resulting eigenvalues, Fig.~\ref{fig:collective_supp}(a), change only weakly over the curvature range of Fig.~3 of the main text, and they are symmetric under $H\to-H$: the intrinsic contributions enter through the Gaussian curvature $K$, which does not distinguish a convex from a concave surface. The asymmetry reported in the main text is therefore driven by $V_H$.

\subsection{Emitters on a spheroid}\label{supp:spheroid_results}
The sphere has $\sigma^{ab}=0$ everywhere and probes $V_H$ alone. To activate the anisotropic operator we deform the interface into a spheroid with equatorial semi-axis $a$ and polar semi-axis $c$, recovering the sphere at $c/a=1$ and the flat interface as $c\to0$. We keep $a=R=62.5\,\bar\lambda_{\rm spp}$ fixed and vary $c/a$ between $0.1$ and $2$, the upper bound respecting the assumption of slowly varying curvature, $\partial_\theta H\ll H$. The pole curvature $H=-sc/a^2$, cf. Eq.~\eqref{eq:H_spheroid}, then stays within $H\bar\lambda_{\rm spp}\in[-0.032,-0.0016]$. All remaining parameters are those of Fig.~3 of the main text.

\begin{figure}[t]
\centering\includegraphics[width=\textwidth]{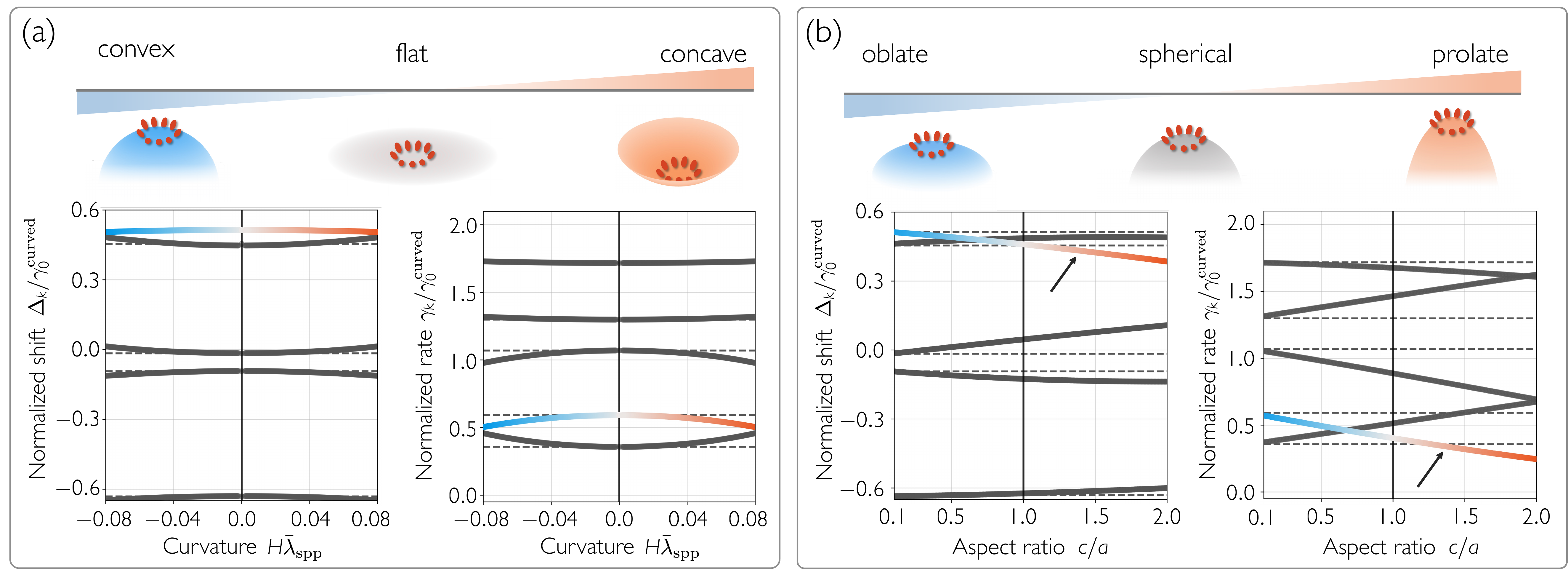}
\caption{\textit{Role of the linear potential and of the spheroidal geometry.} Collective frequency shifts (left) and decay rates (right) for a ring of $N=9$ emitters near the pole of a silver--air interface at $\lambda_0=600\,\mathrm{nm}$, normalized to the single-emitter decay rate $\gamma_0^{\rm curved}$ on the same surface. The nearest-neighbor geodesic spacing is kept at $3\,\bar\lambda_{\rm spp}$ while the geometry is varied, and dashed horizontal lines mark the planar reference values. Modes $k$ and $N-k$ are degenerate, so the nine modes appear as five distinct curves; the full spectrum is shown in dark gray with one eigenmode highlighted by a color gradient. (a)~Sphere, as in Fig.~3 of the main text, but with $V_H$ artificially set to zero, so that all curvature effects arise from the intrinsic geometry contained in $\Delta_\gamma$. The eigenvalues then change only weakly and are symmetric under $H\to-H$. (b)~Spheroid with fixed equatorial semi-axis $a=R=62.5\,\bar\lambda_{\rm spp}$, versus the aspect ratio $c/a$ from oblate ($c/a<1$) through spherical ($c/a=1$, vertical line, corresponding to $H\bar\lambda_{\rm spp}=-0.016$) to prolate ($c/a>1$); arrows mark the highlighted mode. The oblate limit $c\to0$ approaches the flat interface.}
\label{fig:collective_supp}
\end{figure}

Figure~\ref{fig:collective_supp}(b) shows the collective eigenvalues as a function of $c/a$. They evolve asymmetrically about the spherical case: making the interface slightly oblate can raise the radiance of a given mode while making it prolate lowers it, and vice versa. The decay rates respond more strongly than the frequency shifts, and the highlighted mode darkens as the surface becomes prolate, while its cooperative shift changes comparatively little. As on the sphere, modes $k$ and $N-k$ remain degenerate, since a surface of revolution retains the mirror symmetry that exchanges them.

The response to eccentricity is nevertheless dominated by $V_H$ rather than by $V_\sigma$. The pole of a surface of revolution is an umbilic point, at which the two principal curvatures coincide and $\sigma^a{}_b$ vanishes. Away from it,
\begin{equation}\label{eq:umbilic_suppression}
    \frac{\sigma^\theta{}_\theta}{H} = \frac{\left(a^2-c^2\right)\sin^2\theta}{a^2+\rho} \;\xrightarrow[\ \theta\to0\ ]{}\; \frac{1}{2}\left[1-\left(\frac{c}{a}\right)^2\right]\sin^2\theta_0,
\end{equation}
so the anisotropy grows only quadratically in the polar angle of the ring. With a ring radius of $\approx 4.3\,\bar\lambda_{\rm spp}$ at $a=62.5\,\bar\lambda_{\rm spp}$, i.e., $\sin^2\theta_0\approx 5\times10^{-3}$, the ratio in Eq.~\eqref{eq:umbilic_suppression} stays below $1\,\%$ even at $c/a=2$. Since the material prefactors are comparable, e.g., $|C_\sigma k_{\rm spp}^2/C_H|\approx 0.9$ for silver in air, the geometric suppression carries over to the potentials. Figure~\ref{fig:collective_supp}(b) therefore shows essentially the isotropic response of Fig.~3 of the main text, re-parametrized through $H=-sc/a^2$, with the intrinsic contributions bounded by panel~(a). Probing $V_\sigma$ with emitters requires placing them away from umbilic points, for instance near the waist of a catenoid, where the anisotropy is largest.

\section{Emitter decay into SPP modes on flat interfaces}\label{supp:derivation_constant_C0}
\subsection{General equations}
We consider a planar interface at $z=0$ separating a dielectric ($\epsilon_d > 0$, $z>0$) and a metal ($\epsilon_m < 0$, $z<0$) domain. Both media are non-magnetic ($\mu_d = \mu_m = 1$). A quantum emitter modeled as a normally oriented electric dipole $\mathbf{p} = p_z \hat{\mathbf{z}}$ is located at $\mathbf{r}_0 = (0, 0, z_0)$ with $z_0 = d$. The electric field emitted by the dipole is governed by the free-space dyadic Green's tensor. For the vertically oriented dipole, the p-polarized contribution reads~\cite{novotny2012principles}
\begin{equation}
    G_{0,zz}(\mathbf{r}, \mathbf{r}_0) = \frac{i}{8\pi^2 k_d^2} \iint_{-\infty}^{\infty} \frac{k_\parallel^2}{k_{dz}} e^{i [k_x(x-x_0) + k_y(y-y_0) + k_{dz}|z-z_0|]} dk_x dk_y,
\end{equation}
where $k_d = k_0 \sqrt{\epsilon_d}$ is the wavenumber in the dielectric, $k_\parallel^2 = k_x^2 + k_y^2$ is the squared in-plane momentum, and $k_{dz}~=~\sqrt{k_d^2 - k_\parallel^2}$ is the out-of-plane wavenumber, with, importantly, $\mathrm{Im}\left[k_{dz} \right]\ge 0$. The scattered field $G_{zz}^{\rm ref}$, generated by the presence of the metal interface, is found by setting $x_0 = y_0 = 0$, so the reflected tensor component reads
\begin{equation}
    G_{zz}^{\rm ref}(\mathbf{r}, \mathbf{r}_0) = \frac{i}{8\pi^2 k_d^2} \iint_{-\infty}^{\infty} \frac{k_\parallel^2}{k_{dz}} r_p(k_\parallel) e^{i [k_x x + k_y y + k_{dz}(z+z_0)]} dk_x dk_y,
\end{equation}
with the reflection coefficient
\begin{equation}
    r_p(k_\parallel) = \frac{\epsilon_m k_{dz} - \epsilon_d k_{mz}}{\epsilon_m k_{dz} + \epsilon_d k_{mz}},
\end{equation}
and $k_{mz} = \sqrt{k_m^2 - k_\parallel^2}$.
So far, we just followed the equations from Ref.~\cite{novotny2012principles}. We exploit the rotational symmetry by substituting $k_x = k_\parallel \cos \phi_k$ and $k_y = k_\parallel \sin \phi_k$. The spatial coordinates are then defined as $x = \rho \cos \phi_r$ and $y = \rho \sin \phi_r$. The integration measure transforms as $dk_x dk_y = k_\parallel dk_\parallel d\phi_k$, and the phase becomes $k_x x + k_y y = k_\parallel \rho \cos(\phi_k - \phi_r)$. Using
\begin{equation}
    \int_0^{2\pi} e^{i k_\parallel \rho \cos(\phi_k - \phi_r)} d\phi_k = 2\pi J_0(k_\parallel \rho),
\end{equation}
we can reduce the 2D surface integral to a 1D  integral over the radial momentum as
\begin{equation}
    G_{zz}^{\rm ref}(\mathbf{r}, \mathbf{r}_0) = \frac{i}{4\pi k_d^2} \int_0^\infty dk_\parallel \, \frac{k_\parallel^3}{k_{dz}} r_p(k_\parallel) J_0(k_\parallel \rho) e^{i k_{dz}(z+z_0)}.
\end{equation}
We now extend the integration path to the entire real axis $(-\infty, \infty)$ and use the identity $J_0(x) = \frac{1}{2}[H_0^{(1)}(x) + H_0^{(2)}(x)]$. Let us define the integrand as
\begin{equation}
    f(k_\parallel) = \frac{k_\parallel^3}{\kappa_d} r_p(k_\parallel) e^{-\kappa_d(z+z_0)}.
\end{equation}
This function has odd parity, since $f(-k_\parallel) = -f(k_\parallel)$. Using the identity
\begin{equation}
    H^{(2)}_0\left( xe^{-i\pi} \right)=-H^{(1)}_0\left( x \right),
\end{equation}
we can rewrite the integral as
\begin{equation}
G_{zz}^{\rm ref}(\mathbf{r}, \mathbf{r}_0) = \frac{i}{8\pi k_d^2}\int_{-\infty}^\infty \frac{k_\parallel^3}{k_{dz}}\,r_p(k_\parallel)\,H_0^{(1)}(k_\parallel\rho)\,e^{ik_{dz}(z+z_0)}\,dk_\parallel.
\end{equation}
Now, we assume that $\mathrm{Im}\left[k_{\parallel} \right]>0$, and  use that for large arguments $|k_{\parallel}\rho|$, we can write
\begin{equation}
    H^{(1)}_0\left( k_{\parallel}\rho \right)\sim \sqrt{\frac{2}{\pi k_{\parallel}\rho}}e^{i\left( k_{\parallel}\rho-\pi/4\right)}\to 0.
\end{equation}
For large $|k_{\parallel}|$, we can furthermore approximate $k_{dz}= \sqrt{k_d^2 - k_\parallel^2}\approx \pm ik_{\parallel}$, such that
\begin{equation}
    e^{ik_{dz}(z+z_0)}\approx e^{-k_{\parallel}(z+z_0)} \to 0
\end{equation}
for $z+z_0>0$, since $\mathrm{Im}\left[k_{dz} \right]\ge 0$. Therefore, the integrand vanishes for $\mathrm{Im}[k_\parallel]>0$ as $|k_\parallel|\to\infty$, which allows us to close the integration contour $\mathcal{C}$ in the upper half of the complex plane by adding a semicircular arc of radius $R\to\infty$. Since the integrand vanishes on this arc, we have
\begin{equation}
\int_{-\infty}^{\infty} = \oint_{\mathcal{C}} - \int_{\rm arc} = \oint_{\mathcal{C}},
\end{equation}
and using the residue theorem, we get
\begin{equation}
G_{zz}^{\rm ref}(\mathbf{r}, \mathbf{r}_0) = \frac{i}{8\pi k_d^2}\cdot 2\pi i\sum_j\,\mathrm{Res}\left[\frac{k_\parallel^3}{k_{dz}}\,r_p(k_\parallel)\,H_0^{(1)}(k_\parallel\rho)\,e^{ik_{dz}(z+z_0)},\, k_{\parallel,j}\right],
\end{equation}
with the sum running over all poles $k_{\parallel,j}$ enclosed by $\mathcal{C}$. The only pole of the integrand in the upper half-plane originates from the zeros of the denominator of $r_p(k_\parallel)$, i.e., at
\begin{equation}
\epsilon_m\,k_{dz} + \epsilon_d\,k_{mz} = 0,
\end{equation}
which is the SPP dispersion relation satisfied at $k_\parallel = k_{\rm spp}$~\cite{maier2007plasmonics}. For a metal with $\mathrm{Im}[\epsilon_m]>0$, this pole is shifted into the upper half-plane. The residue theorem then gives
\begin{equation}
G_{zz}^{\rm spp}(\mathbf{r}, \mathbf{r}_0) = \frac{1}{k_0^2\epsilon_d} \frac{k_{\rm spp}^3}{\kappa_d}\,\frac{i}{4}H_0^{(1)}(k_{\rm spp}\rho)\,e^{-\kappa_d(z+z_0)}\cdot\mathrm{Res}\left[r_p, k_{\rm spp}\right],
\end{equation}
with $k_{dz}^{\rm spp} = k_{dz}(k_{\rm spp})=i\kappa_d$. Since $r_p$ has a simple pole at $k_{\rm spp}$, the residue is
\begin{equation}
\mathrm{Res}\left[r_p, k_{\rm spp}\right] = \frac{\epsilon_m k_{dz} - \epsilon_d k_{mz}}{\frac{d}{dk_\parallel}\left(\epsilon_m k_{dz} + \epsilon_d k_{mz}\right)}\Bigg|_{k_\parallel=k_{\rm spp}}= -\frac{2k_0}{(\epsilon_m-\epsilon_d)}\left(\frac{\epsilon_d\epsilon
    _m}{\epsilon_m+\epsilon_d} \right)^{3/2}.
\end{equation}
So we can finally write 
\begin{equation}
    G_{zz}^{\rm spp}(\mathbf{r}, \mathbf{r}_0) =C_0\,e^{-2\kappa_d d} \,\, G^{\rm flat}_{\rm 2D}(\mathbf{q}, \mathbf{q}'),
\end{equation}
with the simplified coefficient
\begin{equation}
    C_0 = \frac{-2 k_0 \epsilon_m^3 \epsilon_d \sqrt{-(\epsilon_m + \epsilon_d)}}{(\epsilon_m - \epsilon_d)(\epsilon_m + \epsilon_d)^3}.
\end{equation}
Here, we have set $z=d$, since for the interaction between emitters, we need the evaluation at the positions of the dipoles~\cite{novotny2012principles}.

\subsection{Lossy metals}
Considering material loss, we write $\epsilon_m = \epsilon_m' + i\epsilon_m''$ with $\epsilon_m' < -\epsilon_d < 0$ and $0 < \epsilon_m'' \ll |\epsilon_m'|$, and expand each factor in $C_0$ to first order in $\epsilon_m''$. For the cubic factor in the numerator, we find
\begin{equation}
    \epsilon_m^3 \approx (\epsilon_m')^3\!\left(1 + \frac{3i\epsilon_m''}{\epsilon_m'}\right).
\end{equation}
For the square root, since $\epsilon_m'+\epsilon_d < 0$ we have $-(\epsilon_m+\epsilon_d) = |\epsilon_m'+\epsilon_d| - i\epsilon_m''$, giving
\begin{equation}
    \sqrt{-(\epsilon_m+\epsilon_d)} \approx \sqrt{|\epsilon_m'+\epsilon_d|}\left(1 + \frac{i\epsilon_m''}{2(\epsilon_m'+\epsilon_d)}\right),
\end{equation}
where we used $|\epsilon_m'+\epsilon_d| = -(\epsilon_m'+\epsilon_d)$ to absorb the sign. For the two denominator factors, we find
\begin{equation}
    (\epsilon_m - \epsilon_d) \approx (\epsilon_m'-\epsilon_d)\!\left(1 + \frac{i\epsilon_m''}{\epsilon_m'-\epsilon_d}\right), \qquad (\epsilon_m+\epsilon_d)^3 \approx (\epsilon_m'+\epsilon_d)^3\!\left(1 + \frac{3i\epsilon_m''}{\epsilon_m'+\epsilon_d}\right).
\end{equation}
Writing
\begin{equation}
    C_0 = C_0^{(0)}(1 + i\epsilon_m'' \beta), \qquad 
    C_0^{(0)} = -
    \frac{2k_0(\epsilon_m')^3\epsilon_d\sqrt{|\epsilon_m'+\epsilon_d|}}{(\epsilon_m'-\epsilon_d)(\epsilon_m'+\epsilon_d)^3}\,
    \in \, \mathbb{R},
\end{equation}
and combining the four first-order contributions from numerator and denominator via $(1+ia)(1+ib)^{-1}\approx 1+i(a-b)$, we obtain
\begin{equation}
    \frac{\mathrm{Im}(C_0)}{\mathrm{Re}(C_0)}  = \epsilon_m''\left(\frac{3}{\epsilon_m'} - \frac{5}{2(\epsilon_m'+\epsilon_d)} - \frac{1}{\epsilon_m'-\epsilon_d}\right).
\end{equation}
Using Ref.~\cite{johnson1972optical}, a short calculation shows that for gold at $\lambda_0 = 800\,\mathrm{nm}$ with $\epsilon_m \approx -24.15 + 1.51i$, we get $\mathrm{Im}(C_0)/\mathrm{Re}(C_0) \approx 0.035$, and for silver at $\lambda_0 = 600\,\mathrm{nm}$ with $\epsilon_m \approx -16.12 + 0.44i$, we find $\mathrm{Im}(C_0)/\mathrm{Re}(C_0) \approx  0.016$.

\end{document}